\documentclass[twocolumn,pra,superscriptaddress,showpacs]{revtex4}
\usepackage{color}
\usepackage{amsmath}
\usepackage{amssymb}
\usepackage{graphicx}
\usepackage{esint}
\usepackage{makecell}
\usepackage{array}

\begin{document}

\preprint{APS/123-QED}

\title{Mechanically modulated emission spectra and blockade of polaritons}

\author{Sai-Nan Huai}
\affiliation{Institute of Microelectronics, Tsinghua University, Beijing 100084,
China}
\author{Yu-Long Liu}
\affiliation{Institute of Microelectronics, Tsinghua University, Beijing 100084,
China}

\author{Yunbo Zhang}
\affiliation{Institute of Theoretical Physics, Shanxi University, Taiyuan 030006,
China}

\author{Yu-xi Liu}
\email{yuxiliu@mail.tsinghua.edu.cn}
\affiliation{Institute of Microelectronics, Tsinghua University, Beijing 100084,
China}

\affiliation{Tsinghua National Laboratory for Information Science and Technology
(TNList), Beijing 100084, China}

\date{\today}

\begin{abstract}
We study a hybrid semiconductor-optomechanical system, which consists of a cavity with an oscillating mirror made by semiconducting materials or with a semiconducting membrane inside. The cavity photons and the excitons in the oscillating mirror or semiconducting membrane form into polaritons. And correspondingly, the optomechanical interaction between the cavity photons and the mirror or membrane is changed into the polariton-mechanical interaction. We theoretically study the eigenenergies and eigenfunctions of this tripartite hybrid system with the generalized rotating-wave approximation. We show that the emission spectrum of polariton mode is modulated by the mechanical resonator. We also study the mechanical effect on the statistical properties of the polariton when the cavity is driven by a weak classical field. This work provides a detailed description of the rich nonlinearity owing to the competition between parametric coupling and three-wave mixing interaction concerning the polariton modes and the phonon mode. It also offers a way to operate the photons, phonons and excitons, e.g., detect the properties of mechanical resonator through the fine spectra of the polaritons or control the transmission of light in the integrated semiconducting-optomechanical platform.

\end{abstract}

\pacs{Valid PACS appear here}% PACS, the Physics and Astronomy
\maketitle

%\tableofcontents

\section{\label{sec:level1}INTRODUCTION}

Cavity optomechanical systems, which consist of single-mode cavity fields and mechanical resonators, have attracted growing interest for its potential applications in ultrasensitive force sensors, frequency conversion, high-precision measurements, and quantum information processing~\cite{kippenberg,Thompson2008,marquart,aspelmeyer,Revmodphys,Liu2018}. The masses of mechanical resonators in optomechanical systems vary from picograms to kilograms, meanwhile their frequencies usually range from hundreds of megahertz down to the hertz level. Although the frequencies of cavity fields in most experimental studies for optomechanical systems vary from optical domain to microwave and radiowave domain, electromagnetic fields with any wavelengths can still couple to the mechanical resonator. In cavity optomechanics, the coupling between cavity field and mechanical oscillator can have different mechanisms. The most common ones are radiation pressure force or photothermal force, both of which originate from momentum transfer due to reflection or absorption of photons.

Various materials are used to construct optomechanical systems in order to increase or control optomechnaical coupling, but the detailed properties of the materials themselves are usually less considered. For example, there are studies that optomechanical systems are coupled to either two-level or other systems via either cavity fields~\cite{Ian2008,Chang2011,Jing2011,Wang2012,Jing2012} or mechanical resonators~\cite{Tian2011,Ramos2013,Wang2014,Wang2015}. However, material properties of cavities and mechanical resonators in these studies are not studied. It is known that many optomechanical systems are made of semiconducting materials, for example, cavity optomechanics was demonstrated in gallium arsenide~\cite{Ding2010,Liu2011,Watanabe2012,Usami2012,Xuereb2012,Gil-Santos2017} and gallium phosphide microdisks~\cite{Mitchell2014}. Recently, there are reports on the coupling between mechanical resonator and exciton, which are electron-hole pairs, in GaAs/AlGaAs quantum dot system~\cite{Yeo2014,Montinaro2014}. Also cavity-less optomechanics is demonstrated  by virtue of opto-piezoelectric backaction through excitons in an n-GaAs/i-GaAs bilayer cantilevers~\cite{Okamoto2011, Okamoto2011-PRB}. Such carrier mediated optomechanical coupling does not require any optical cavities but is based on the piezoelectric effect. All of these studies open up a new way to operate electrons, photons and phonons in an integrated semiconducting platform by using semiconducting microcavity quantum electrodynamics (QED) and optomechanics. For example, the spectrum of mechanical oscillation is proposed to detect the fine energy structure of the excitons~\cite{Okamoto2011, Okamoto2011-PRB} in semiconducting materials.

Semiconducting microcavity QED is extensively studied since the observation on the strong coupling between a single-mode cavity field and excitons (electron-hole pairs) in semiconducting quantum well, which is embedded in a microcavity~\cite{Weisbuch1992}. It is well known that the strong coupling between the excitons and photons can mix them and result in so-called polaritons~\cite{Hopfield1958}. In low dimensional semiconductor~\cite{Cao2000,Cao2001,Jin2003} or semiconducting cavity QED~\cite{Savona1999,Ciuti2003} system, the polaritions can be observed through photoluminescene, photon reflection or transmission. Recently, an optomechanical experiment showed that the mechanical modes of a GaAs nano-membrane can be cooled down via photonthermal effect mediated by excitons inside the membrane~\cite{Usami2012,Xuereb2012}. The strong optomechanical coupling was observed through cavity polaritons~\cite{Rozas2014}. Also polariton resonances for ultrastrong coupling cavity optomechanics in GaAs/AlAs quantum wells were demonstrated~\cite{Jusserand2015}. Optomechanics via cavity polaritons~\cite{Kyriienko2014} and exciton-phonon entanglement~\cite{Sete2015} were proposed.

\begin{figure}[ptb]
\includegraphics[scale=0.245,clip]{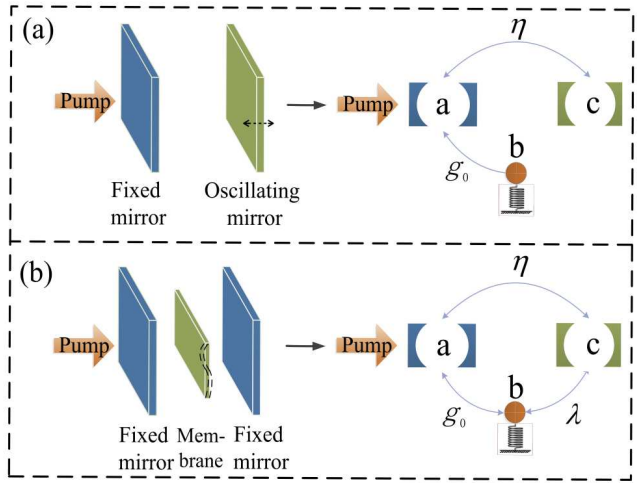}\
%bb=20 220 550 610,  width=0.45\textwidth, 
%\includegraphics[scale=0.25,clip]{fig1b.jpeg}%bb=25 220 573 640, scale=0.5,
 \caption{(Color online) Schematic diagram of a hybrid semiconducting cavity optomechanical system with (a) an oscillating mirror made by the semiconducting materials or (b) a thin semiconducting membrane inside the cavity. In each panel, the right part further shows the coupling relation between different parts of the hybrid system. The parameters $\eta$, $g_{0}$ and $\lambda$ represent the cavity photon-exciton, photon-phonon and exciton-phonon coupling strengths, respectively. We assume that there is no direct coupling between the exciton and mechanical resonator in (a).}%
\label{Figure1}%
\end{figure}

Motivated by recent works~\cite{Usami2012,Xuereb2012,Okamoto2011, Okamoto2011-PRB,Rozas2014,Jusserand2015,Kyriienko2014}, here we study a system that consists of a cavity with an oscillating mirror made by semiconducting materials or with a thin semiconducting membrane inside the cavity. We will show how the mechanical motion affects the emission spectra and blockade of polaritions. The paper is organized as follows. In Sec.~\ref{Sec.II}, we will give a theoretical model to describe the interaction between the exciton, a single-mode cavity field, and the mechanical resonator. Particularly, in Sec.~\ref{Sec.IIA}, we introduce the polariton modes formed by the cavity photons and excitons. In Sec.~\ref{Sec.IIB}, we first present general method to get the eigenvalues of the hybrid system, and then as an example, we study analytically the eigenenergies and eigenfunctions in one polariton subspace of the fully coupled hybrid system by diagonalizing the system Hamiltonian with the generalized rotating-wave approximation (GRWA) proposed in Ref.~\cite{Irish2007}. In Sec.~\ref{Sec.III}, we present our results on the properties of the emission spectra of the polaritons. In Sec.~\ref{Sec.IV}, the statistic properties of the polariton are investigated via the equal-time second-order correlation function and the polariton blockade and tunneling are discussed. Finally, we summarize the results in Sec.~\ref{Sec.V}. The eigenenergies and eigenfunctions in the two-polartion subspace are shown in the Appendix. 

\section{MODEL}\label{Sec.II}
As schematically shown in the left parts of Fig.~\ref{Figure1}, we study a system that consists of a cavity either with an oscillating mirror made of semiconducting materials in Fig.~\ref{Figure1}(a) or a thin semiconducting membrane placed in the middle of the cavity in Fig.~\ref{Figure1}(b). Besides, in the right parts of Figs.~\ref{Figure1} (a) and (b), we show the coupling relationship corresponding to the left ones. The difference of these two situations is whether there is direct coupling between the excitons in the semiconductor and the mechanical resonator, while we assume that the couplings of the photons to excitons and phonons exist in both configurations. The Hamiltonian of the whole system can be written as $(\hbar=1)$
\begin{eqnarray}
H&=&\omega_{c} a^{\dagger}a+\omega_{m} b^{\dagger}b+\omega_{\rm ex} c^{\dagger}c\nonumber\\
&&+g_{0}a^{\dagger }a(b^{\dagger }+b)+\lambda c^{\dagger
}c\left( b^{\dagger }+b\right) \nonumber\\
&&+\eta \left( a^{\dagger }c+c^{\dagger
}a\right). \label{H}
\end{eqnarray}

The first line in Eq.~(\ref{H}) is the free Hamiltonian of the system. The operators $a\ (a^{\dagger})$, $b\ (b^{\dagger})$ and $c\ (c^{\dagger})$ are, the annihilation (creation) operators of the cavity field, mechanical resonator and exciton, with corresponding resonant frequencies  $\omega_{c}$, $\omega_{m} $ and $\omega_{ex}$, respectively. The rest terms in Eq.~(\ref{H}) describe the interaction in the fully coupled tripartite hybrid system. The parameter $g_{0}=\omega_{c}x_{zpf}/L$ represents the single photon-phonon coupling caused by the radiation pressure between the cavity and the mirror, where $x_{zpf}$ is the mechanical zero-point fluctuations and $L$ is the length of the cavity. The parameter $\lambda$ denotes the deformation-potential coupling between the exciton and mechanical resonator, while $\eta$ describes  the interaction between the cavity field and the exciton with rotating wave approximation. For the case shown in Fig.~\ref{Figure1}(a), in which the oscillating mirror made of semiconducting materials moving in translation but without deformation, we assume that there is no direct coupling between the exciton and the phonon, i.e., $\lambda=0$. In our paper, we will mainly study the model shown in Fig.~\ref{Figure1}(b) for generality. Our research results can be applied to the case shown in Fig.~\ref{Figure1}(a) by setting $\lambda=0$.

\subsection{Polariton modes} \label{Sec.IIA} 
When the photons are coupled to the excitons, the cavity and exciton modes are hybridized into polariton modes. The polariton modes can be expressed as~\cite{Hopfield1958}
\[\left(\begin{array}{c}
A \\ 
B%
\end{array}%
\right) =\left( 
\begin{array}{cc}
\cos \theta & \sin \theta \\ 
-\sin \theta & \cos \theta%
\end{array}%
\right) \binom{a}{c}, 
\]\\
where  $\tan 2\theta =2\eta/\Delta_{ce}$ and $\theta \in \left[ 0,\pi/2\right]$, with $\Delta_{ce}=\omega_{ex}-\omega_{c}$ the detuning between the exciton and the photon. This transformation shows that for different detuning $\Delta_{ce}$ and coupling strength $\eta$ between the photons and the excitons, the photonic and excitonic components vary a lot in polariton modes A and B. When the detuning $\Delta_{ce}$ is very large, the polariton mode A approaches either bare cavity (positive infinity detuning) or exciton mode (negative infinity detuning), and vice versa for mode B.  Besides, for the resonant case, i.e, $\theta=\pi/4$, we can get the maximally hybridized polariton modes $A/B=(c\pm a)/\sqrt2$, where $A$ ($B$) mode corresponds to the sign $+$ ($-$). Using the polariton modes, we can rewrite the total Hamiltonian in Eq.~(\ref{H}) as
\begin{eqnarray}
H_{1}&=&\omega _{A}A^{\dagger }A+\omega _{B}B^{\dagger
}B+\omega _{m}b^{\dagger }b \nonumber\\
&&+ \left (Q_{A}A^{\dagger }A+Q_{B}B^{\dagger }B \right)\left( b^{\dagger }+b\right)\nonumber\\
&&+ Q\left( A^{\dagger }B +B^{\dagger }A\right)\left( b^{\dagger }+b\right),\label{H1}
\end{eqnarray}
where $\omega_{A}$ and $\omega_{B}$ are the eigenfrequencies of the two polariton modes with  
\begin{equation}
\omega _{A/B} =\frac{1}{2}\left(\omega _{a}+\omega _{ex}\right)\pm \frac{\eta }{\sin2\theta },
\end{equation}
where the sign $+$ ($-$) corresponds to $\omega_A$ ($\omega_B$). Parameters $Q_{A}, Q_{B}$ ($Q$) are the dispersive (three-wave mixing interaction) polariton-phonon coupling strengths. They can be expressed in terms of the original parameters as
\begin{eqnarray}
Q_{A}&=g_{0}\cos^{2}\theta +\lambda\sin^{2}\theta, \label{QA}\\
Q_{B}&=g_{0}\sin^{2}\theta +\lambda\cos^{2}\theta, \label{QB}\\
Q&=\left(\lambda-g_{0}\right)\cos\theta\sin\theta.\label{Q}
\end{eqnarray}
One can find that the coupling strengths between the polaritons and mechanical mode are modified, since the polariton modes share contributions from both cavity photons and excitons.  
\subsection{Eigenenergies and eigenstates} \label{Sec.IIB}
 
To get further insight into the nature of the fully coupled tripartite system, we now study the eigenenergies and eigenstates of the Hamiltonian in Eq.~(\ref{H1}). We find that it is convenient to obtain the solutions using Schwinger's representation of the augular momentum for the two bosonic polariton modes $A$ and $B$. Using the polariton operators, angular-momentum operators can be constructed as~\cite{Beidenharn1981}
\begin{eqnarray}
J_{x} =\frac{1}{2}\left(A^{\dagger }B+B^{\dagger }A\right)&,& J_{y}=\frac{1}{2i}\left(A^{\dagger }B-B^{\dagger }A\right),\nonumber\\
J_{z}&=&\frac{1}{2}\left(A^{\dagger}A-B^{\dagger }B\right).
\end{eqnarray}
The total anugular-momentum operator is given as
\begin{equation}
J^{2}=J_{x}^{2}+J_{y}^{2}+J_{z}^{2}=\frac{{N}}{2}\left(\frac{{N}}{2}+1\right),
\end{equation}
where ${N}=A^{\dagger }A+B^{\dagger }B$ denotes the total polariton number operator of modes $A$ and $B$. For given total excitation number $\mathcal{N}$ of polaritons , the simultaneous eigenstates of $J^{2}$ and $J_{z}$ are defined as
\begin{equation}
\left\vert j,m\right\rangle=\frac{\left(A^{\dagger}\right)^{j+m}\left(B^{\dagger}\right)^{j-m}}{\sqrt{(j+m)!(j-m)!}}\left\vert 0\right\rangle,
\end{equation}
with the eigenvalues $j=\mathcal{N}/2$, and $m=-\mathcal{N}/2,...,\mathcal{N}/2$. In order to get more intuitive understanding, we can also expressed $\left\vert j,m\right\rangle$ as $\left\vert n_{A}\right\rangle\otimes\left\vert n_{B}\right\rangle$ in terms of polariton Fock state. That is, $\left\vert j,m\right\rangle$ =$\left\vert n_{A}\right\rangle\left\vert n_{B}\right\rangle$, where 
\begin{equation}
n_{A}=j+m,\text{ \ \ \ } n_{B}=j-m \label{NANB}
\end{equation}
represent the excitation numbers in modes $A$ and $B$, respectively. Furthermore, it can be verified that Eq.~(\ref{H1}) can be transformed into~\cite{Liu2001, Liu2002} 
 \begin{align}
H_{2} =&\frac{1}{2}\left(\omega _{A}+\omega _{B}\right){N}+\left( \omega _{A}-\omega_{B}\right) \left(\cos\phi J_{z}+\sin\phi J_{x}\right)\nonumber\\
&+\omega _{m}b^{\dagger }b+\left( \Omega {N}+GJ_{z}\right)\left( b^{\dagger }+b\right)  \label{H2}
\end{align}
by performing a rotation $U_{1}=$exp$\left({-i\phi J_{y}}\right)$ with $\phi=2\theta$. And other parameters are given as
\begin{eqnarray}
 \Omega=\frac{1}{2}(Q_{A}+Q_{B}),\text{ \ \ \ } G=g_0-\lambda.\label{EqT}
\end{eqnarray}
In order to tackle the static shift of the mechanical resonator equilibrium position caused by polartions, we implement another unitary transform
\begin{equation}
U_{2}=\exp\left[\left(\Omega {N}+GJ_{z}\right)\left( b^{\dagger }-b\right)/\omega _{m}\right]
\end{equation}
to Eq.~(\ref{H2}). This transformation displaces the creation and annihilation operators of the mechanical resonator by $\left(\Omega {N}+GJ_{z}\right)/\omega_{m}$. Then we obtain an effective Hamiltonian $H_{3}=U_{2}H_{2}U_{2}^{\dagger}$ with
\begin{align}
H_{3}=&\frac{1}{2}\left(\omega _{A}+\omega _{B}\right){N}+\left(\omega _{A}-\omega _{B}\right)\cos\phi J_{z}+\omega_{m}b^{\dagger}b \nonumber\\
&+\left(\omega _{A}-\omega _{B}\right)\sin\phi J_{x}\cosh\left[ \frac{G}{\omega_m}\left(b^{\dagger}-b \right) \right] \nonumber\\
&+i\left(\omega _{A}-\omega _{B}\right)\sin\phi J_{y}\sinh\left[ \frac{G}{\omega_m}\left(b^{\dagger}-b \right) \right]      \nonumber\\ 
 &-\frac{1}{\omega_{m}}\left(\Omega {N}+GJ_{z}\right)^{2}. \label{H3}
\end{align}

For large photon-exciton detuning ($\Delta_{ce}=\omega_{ex}-\omega_{c}\gg\eta$ so that $\phi=0$) or the coupling balanced case ($g_0=\lambda$), the terms in the second and third lines of Eq.~(\ref{H3}) vanish, which means the mechanical resonator $b$ is decoupled from the polariton modes $A$ and $B$. In these cases, the Hamiltonian in Eq.~(\ref{H3}) is reduced to 
\begin{align}
\tilde{H}_{3}=&\frac{1}{2}\left(\omega _{A}+\omega _{B}\right){N}+\left(\omega _{A}-\omega _{B}\right)J_{z}+\omega_{m}b^{\dagger}b \nonumber\\
&-\frac{1}{\omega_{m}}\left(\Omega {N}+GJ_{z}\right)^{2}.
\end{align}
Then the eigenengergies of the original Hamiltonian in Eq.~(\ref{H1}) can be easily obtained as
\begin{align}
E_{j,m,n_{b}}=&j\left(\omega_{A}+\omega_{B}\right)+m\left(\omega_{A}-\omega_{B}\right)+n_{b}\omega_{m}\nonumber\\
&-(j+m)^{2}\Delta_{A0}-(j-m)^{2}\Delta_{B0}\nonumber\\
&-2(j+m)(j-m)\Delta_{AB0}.  \label{Energy}
\end{align}
Here $n_b$ denotes the phonon excitation number corresponding to the phonon number operator $N_b=b^{\dagger}b$. The parameters $\Delta_{A0}={g_{0}^{2}}/{\omega_m}$, $\Delta_{B0}={\lambda^{2}}/{\omega_m}$, and $\Delta_{AB0}={g_{0}\lambda}/{\omega_m}$ describe the frequency shifts and the nonlinearity of the polariton modes, caused by the dispersive coupling to the phonon mode $b$. The corresponding eigenfunctions can be given by
\begin{equation}
\psi_{j,m,n_{b}}=\left\vert j,m\right\rangle\left\vert{n}_{b}\right\rangle_{{j,m}}.
%\equiv\left\vert j,m,n_b\right\rangle .
\end{equation}
Here, the state 
\begin{equation}
\left\vert{n}_{b}\right\rangle_{{j,m}}=e^{-\beta_{j,m}\left(b^{\dagger}-b\right)}\left\vert n_{b}\right\rangle
\end{equation}
is a (j,m)-polaritons displaced Fock state~\cite{Oliveira1990}, where $\beta_{j,m}=\left(\mathcal{N}\Omega+mG\right)/\omega_{m}$ denotes the displacement determined by the angular momentum number $\left(j,m\right)$. 

However, for the most common case, the three-wave mixing interaction included in the second and third lines of Eq.~(\ref{H3}) also play an important role and will surely induce more nonlinear terms. We can expand the hyperbolic functions $\cosh$ and $\sinh$, respectively, as 
\begin{eqnarray}
\cosh\left[ \frac{G}{\omega_m}\left(b^{\dagger}-b \right)\right]&=&G_0(b^{\dagger}b)+G_1(b^{\dagger}b){b^{\dagger}}^{2} \nonumber\\
 &&+b^{2}G_{1}(b^{\dagger}b)+\cdots,\\
\sinh\left[\frac{G}{\omega_m}\left(b^{\dagger}-b \right)\right]&=&F_1(b^{\dagger}b)b^{\dagger}-bF_1(b^{\dagger}b)\nonumber\\
&&+F_2(b^{\dagger}b){b^{\dagger}}^{3}-b^{3}F_{2}(b^{\dagger}b) \nonumber\\
&&+\cdots. 
\end{eqnarray}
Here, $G_i(b^{\dagger}b)(i=0,1,\cdots)$ and $F_j(b^{\dagger}b)(j=1,2,\cdots)$ are coefficients that depend on the phonon number operator $N_b=b^{\dagger}b$ and the dimensionless parameter $G/\omega_m$. Different orders of approximations can then be performed by only keeping some primary terms while neglecting others in the expansions.

\begin{figure}[ptb]
\includegraphics[scale=0.35,clip]{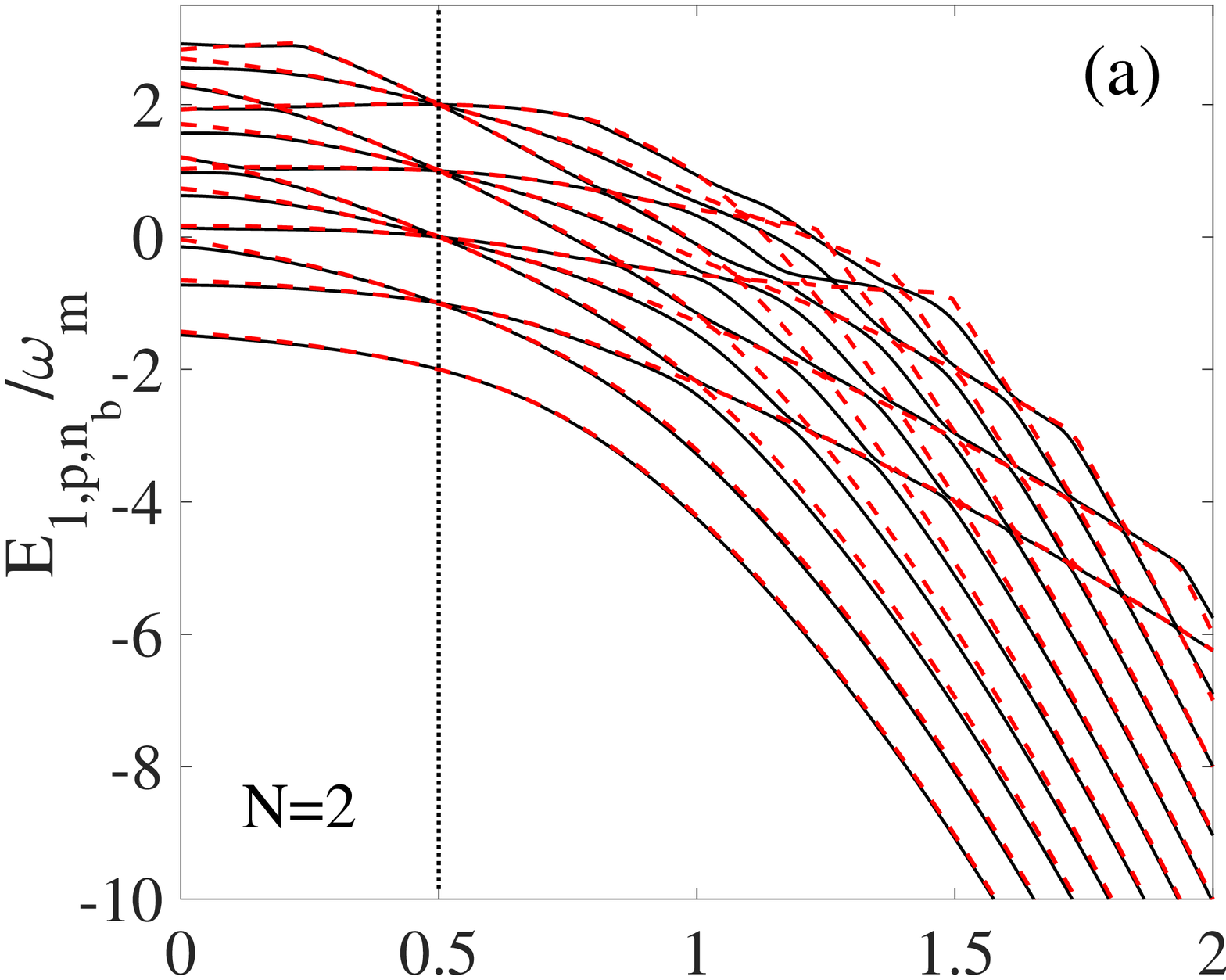}\
\includegraphics[scale=0.35,clip]{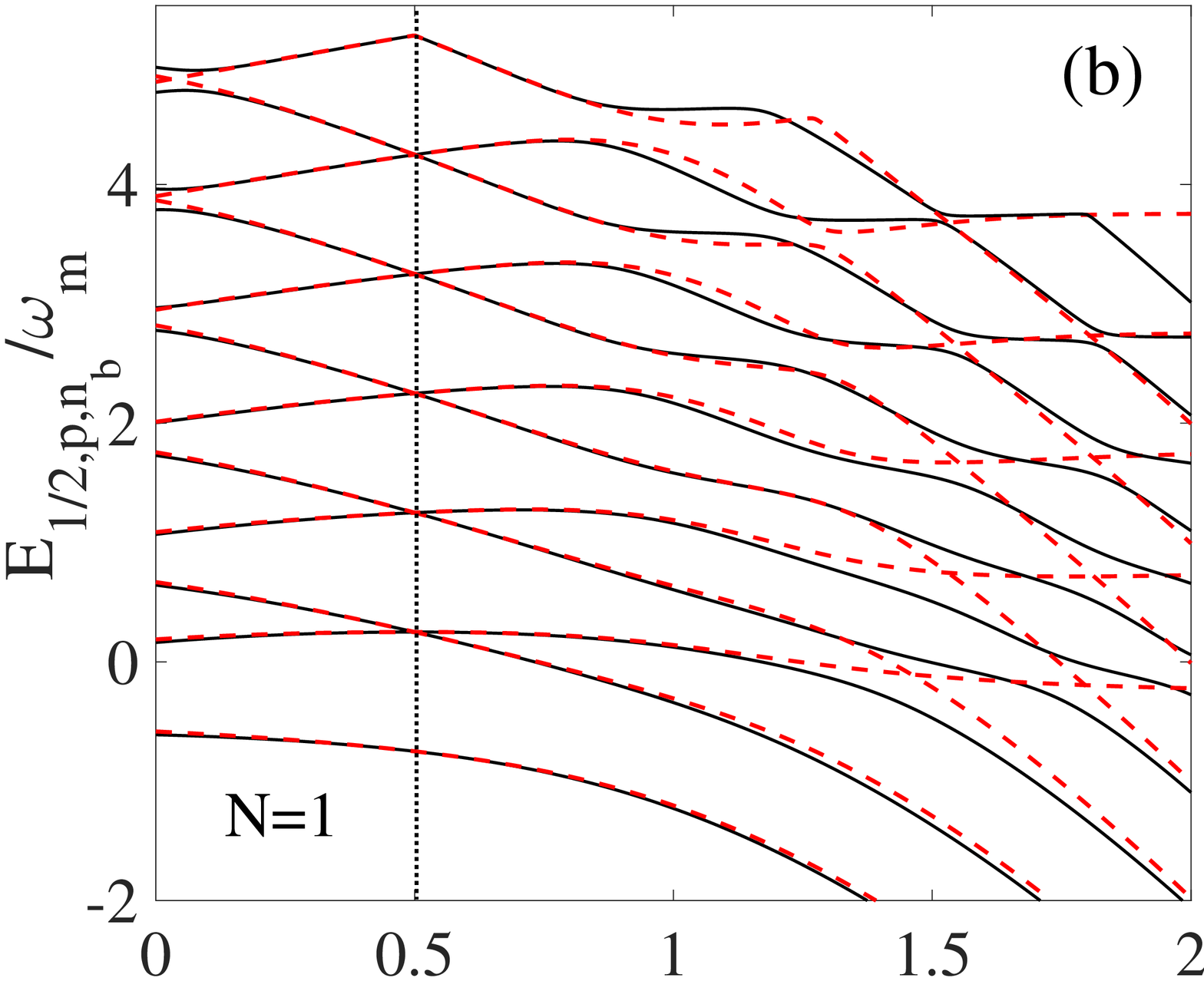}\
\includegraphics[scale=0.35,clip]{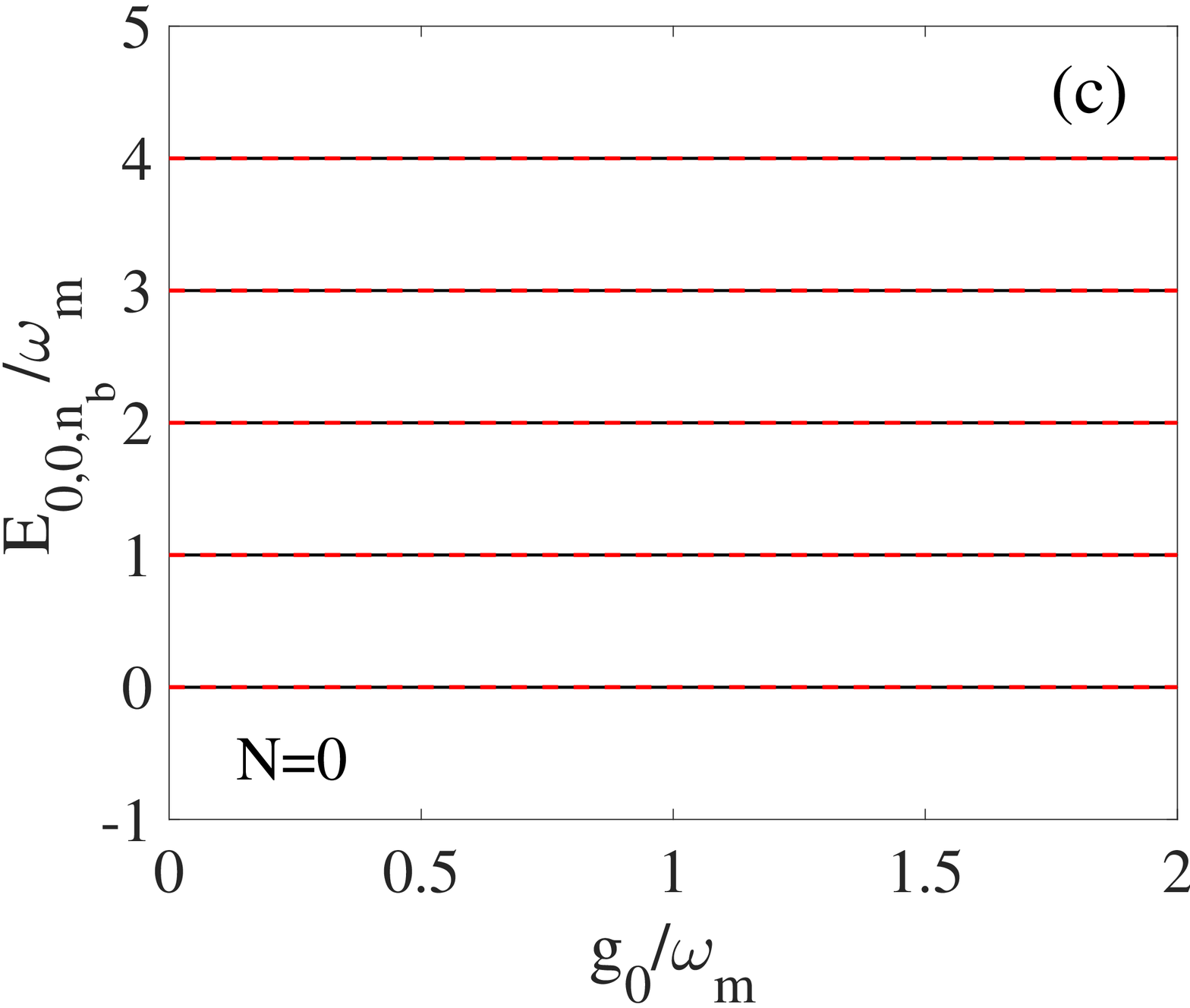}\
%bb=20 220 550 610,  width=0.45\textwidth, 
%\includegraphics[scale=0.25,clip]{fig1b.jpeg}%bb=25 220 573 640, scale=0.5,
 \caption{(Color online) Energy levels of the system for polariton excitation numbers (a) $\mathcal{N}=2$, (b) $\mathcal{N}=1$, (c) $\mathcal{N}=0$ versus the coupling strength $g_{0}/\omega_{m}$ when cavity photons and excitons are resonant, i.e., $\Delta_{ce}=0$. The black-solid curves represent the numerical result while the red-dotted curves are following the GRWA result Eq.~(\ref{EGRWA}). Other parameters are: $\eta$=0.5 $\omega_m$, $\lambda$=0.5 $\omega_m$. }%
\label{Figure2}%
\end{figure}

We note that the total polariton number operator ${N}$ commutes with the total Hamiltonian of the system, i.e., $\left[H_{3}, {N}\right]=0$, thus the Hamiltonian of the closed system can be block-diagonalized in the basis of the eigenvectors of the polariton number operator. When there is no polariton excitation, i.e, $\mathcal{N}=0$, the eigenenergy behaves just like the usual harmonic structure. As shown in Fig.~\ref{Figure2}(c), it reveals that the eigenenergies are independent of the coupling strength $g_0$ between the cavity field and mechanical resonator. For the $\mathcal{N}=1$ subspace, we first consider the zeroth-order approximation which neglects the terms involving energy exchange between the phonon and the polaritons. In this case, $J_x, J_y$, and $J_z$ are defined in two dimensional space and equivalent to Pauli operators, i.e., $J_x=\sigma_x/2,J_y=\sigma_y/2$, and $J_z=\sigma_z/2$. Then the Hamiltonian in Eq.~(\ref{H3}) can be approximated as 
\begin{eqnarray}
H_{3}^{\left( 0\right) }&=&\frac{1}{2}\left(\omega _{A}-\omega _{B}\right)\cos
\phi \sigma _{z}+\omega _{m}b^{\dagger }b-\frac{1}{\omega _{m}}\left( \Omega +\frac{G}{2}%
\sigma _{z}\right) ^{2} \nonumber \\
&&+\frac{1}{2}\left(\omega _{A}-\omega _{B}\right)\sin
\phi \sigma _{x}G_{0}\left( b^{\dagger }b\right).\label{H30}
\end{eqnarray}
Note that for the sake of clarity we have neglected the energy baseline $\left(\omega _{A}+\omega _{B}\right)/2$. And there are only terms concerning the phonon number operator $N_b=b^{\dagger}b$. Thus logically, the Hilbert space can be decomposed into $n_b$ manifolds in the basis of the angular momentum and mechanical resonator $ \left\vert \frac{1}{2},-\frac{1}{2},n_b\right\rangle$ and $\left\vert \frac{1}{2},\frac{1}{2},n_b\right\rangle$. Based on the fact that 
\begin{eqnarray*}
\sigma _{x}\left\vert \frac{1}{2},\frac{1}{2}\right\rangle &=&\left\vert
\frac{1}{2},-\frac{1}{2}\right\rangle ,\text{ \ \ \ \ \ \ \ \ \ }\sigma
_{x}\left\vert \frac{1}{2},-\frac{1}{2}\right\rangle =\left\vert \frac{1}{2},%
\frac{1}{2}\right\rangle, \\
\sigma _{y}\left\vert \frac{1}{2},\frac{1}{2}\right\rangle &=&i\left\vert 
\frac{1}{2},-\frac{1}{2}\right\rangle ,\text{ \ \ \ \ \ \ \ }\sigma
_{y}\left\vert \frac{1}{2},-\frac{1}{2}\right\rangle =-i\left\vert \frac{1}{2%
},\frac{1}{2}\right\rangle,
\end{eqnarray*}%
the Hamiltonian in Eq.~(\ref{H30}) in the $n_b$-th manifold takes on the form 
\begin{equation}
H_{3,n_{b}}^{\left( 0\right) }=\left[ 
\begin{array}{cc}
 e_{n_b}^{\left(1\right)}& \frac{B_{n_{b}}}{2} \\ 
\frac{B_{n_{b}}}{2} & e_{n_b}^{\left(2\right)}%
\end{array}%
\right], \label{H3nb0}
\end{equation}
with
\begin{eqnarray}
e_{n_b}^{\left(1\right)}&=&-\frac{1}{2}\left(\omega _{A}-\omega _{B}\right)\cos \phi +n_{b}\omega _{m}-\frac{\lambda^{2}}{\omega _{m}},\\
e_{n_b}^{\left(2\right)}&=&\frac{1}{2}\left(\omega _{A}-\omega _{B}\right)\cos \phi
+n_{b}\omega _{m}-\frac{g_{0}^{2}}{\omega _{m}},\\
B_{n_{b}}&=&\left( \omega _{A}-\omega _{B}\right) \sin \phi G_{0}\left( n_{b}\right).
\end{eqnarray}
Here, for a given phonon number $n_b$, the coefficient $G_0(b^{\dagger }b)$ is given as
 \begin{equation}
G_{0}\left( n_{b}\right)=\exp\left(-\frac{G^2}{2\omega _{m}^{2}}%
\right)L_{n_{b}}\left( \frac{G^2}{\omega _{m}^{2}}%
\right), 
\end{equation}
with the Laguerre polynomials 
\begin{equation}
L_{n}^{m-n}\left( x\right)=\sum_{l=0}^{\min (m,n)}\left( -1\right) ^{n-l}\frac{m!x^{n-l}}{\left( m-l\right) !\left( n-l\right) !l!}.
\end{equation}
Using Eq.~(\ref{H3nb0}), the eigenenergies corresponding to the Hamiltonian in Eq.~(\ref{H30}) can be straightforwardly given by 
\begin{eqnarray}
\varepsilon _{\frac{1}{2},p,n_{b}}&=&n_{b}\omega _{m}-\frac{1}{2\omega _{m}}\left(g_{0}^{2}+\lambda^{2}\right)\nonumber\\
&&\pm \frac{1}{2}\sqrt{\left( e_{n_b}^{(2)}-e_{n_b}^{(1)}\right) ^{2}+B_{n_{b}}^{2}} ,
\end{eqnarray}
where $p=+$ or $p=-$ denotes the two eigenvalues in the subspace of one polariton and $n_b$ phonon excitations. The corresponding eigenfunctions are 
\begin{equation}
\left\vert \varepsilon _{\frac{1}{2},p,n_{b}}\right\rangle=\frac{1}{%
\lambda _{\frac{1}{2},p,n_{b}}}\binom{1}{\mu _{\frac{1}{2},p,n_{b}}} \\
\end{equation}
with
\begin{eqnarray*}
\mu _{\frac{1}{2},p,n_{b}} &=&{\frac{2}{B_{n_{b}}}}{\left(\varepsilon _{\frac{1}{2}%
,p,n_{b}}-e_{n_b}^{(1)}\right) }, \\
\lambda _{\frac{1}{2},p,n_{b}} &=&\sqrt{1+\mu _{\frac{1}{2},p,n_{b}}^{2}}.
\end{eqnarray*}

The validity of the zeroth-order approximation is restricted to the large detuning regime, that is, $\left( \omega_A-\omega_B\right)\cos\phi\ll\omega_m$~\cite{Ashhab2010,Agarwal2012,Mao2015,Mao2016}. However, for the resonant case, i.e., $\left( \omega_A-\omega_B\right)\cos\phi=\omega_m$, the transitions between different phonon number manifolds should be included~\cite{Irish2007,Zhang2015,Zhang2016}. For example, we take the single-phonon exchange terms in the expansion of the hyperbolic functions of Eq.~(\ref{H3}) into account in the first-order approximation. That is, only single phonon exchange between the polaritons and mechanical resonator is considered. 

Now the Hamiltonian in Eq.~(\ref{H3}) can be approximately written into two parts
\begin{eqnarray}
H_{3}^{\left( 1\right) }=H_{3,0}^{\left( 1\right) }+H_{3,1}^{\left( 1\right) },
\end{eqnarray}
with
\begin{eqnarray}
H_{3,0}^{\left( 1\right) }=H_{3}^{\left( 0\right) }-\frac{1}{2}\left(\omega _{A}-\omega _{B}\right)\sin \phi %
 \sigma _{x}\left[G_{0}\left( b^{\dagger }b\right) -\beta \right], 
%\frac{1}{2}\left(\omega _{A}-\omega _{B}\right)\cos \phi \sigma _{z}+\omega
%_{m}b^{\dagger }b-\frac{1}{%
%\omega _{m}}\left( \Omega +\frac{G}{2}\sigma _{z}\right) ^{2} 
\end{eqnarray}
and
\begin{eqnarray}
H_{3,1}^{\left( 1\right) } &=&\frac{1}{2}\left(\omega _{A}-\omega _{B}\right)\sin \phi %
 \sigma _{x}\left[G_{0}\left( b^{\dagger }b\right) -\beta \right]  \nonumber\\
&&+\frac{i}{2}\left(\omega _{A}-\omega _{B}\right)\sin\phi\sigma _{y} \left[ F_{1}\left( b^{\dagger }b\right)b^{\dagger }-bF_{1}\left( b^{\dagger }b\right)\right],\nonumber\\
\end{eqnarray}
where $\beta =G_{0}\left( 0\right) =\exp\left[-{G^{2}}/{2\omega _{m}^{2}}\right]$. Note in $H_{3,0}^{\left( 1\right) }$, the angular momentum and mechanical resonator operators are completely decoupled by applying an unitary transformation 
\begin{eqnarray}
U_3 &=&\left[ 
\begin{array}{cc}
\frac{1}{\lambda _{-}} & \frac{\mu _{-}}{\lambda _{-}} \\ 
\frac{1}{\lambda _{+}} & \frac{\mu _{+}}{\lambda _{+}}%
\end{array}%
\right] 
\end{eqnarray}
to the angular momentum part, and $H_{3,0}^{\left( 1\right) }$ can be diagonalized into
\begin{eqnarray}
\tilde{H}_{3,0}^{\left( 1\right) } =U_{3}H_{3,0}^{\left( 1\right) }U_{3}^{\dagger }=\omega _{m}b^{\dagger }b+\left[ 
\begin{array}{cc}
\varepsilon _{-} & 0 \\ 
0 &\varepsilon _{+}%
\end{array}%
\right], 
\end{eqnarray}
where $\lambda _{\pm }=\lambda _{\frac{1}{2},\pm ,0}$, $\mu _{\pm }=\mu _{\frac{1}{2},\pm ,0}$, and $\varepsilon _{\pm }=\varepsilon _{\frac{1}{2},\pm ,0}$. In this way, the second part $H_{3,1}^{\left( 1\right) }$ of the Hamiltonian $H_{3}^{\left( 1\right) }$ is transformed into:
\begin{eqnarray}
\tilde{H}_{3,1}^{\left( 1\right) } &=&U_{3}H_{3,1}^{\left( 1\right) }U_{3}^{\dagger }\nonumber\\
&=&\frac{1}{2}L\left(\omega _{A}-\omega _{B}\right) \sin \phi \left[
G_{0}\left( b^{\dagger }b\right) -\beta \right]\nonumber\\
&&+\frac{1}{2}M\left(\omega _{A}-\omega _{B}\right) \sin \phi
\left[ F_{1}\left( b^{\dagger }b\right)b^{\dagger }-bF_{1}\left( b^{\dagger }b\right)\right],\nonumber\\
\end{eqnarray}
with
\begin{eqnarray}
L=\left[ 
\begin{array}{cc}
\frac{2\mu _{-}}{\lambda _{-}^{2}} & \frac{\mu _{-}+\mu _{+}}{\lambda
_{-}\lambda _{+}} \\ 
\frac{\mu _{+}+\mu _{-}}{\lambda _{-}\lambda _{+}} & \frac{2\mu _{+}}{%
\lambda _{+}^{2}}%
\end{array}%
\right], \ 
M=\left[ 
\begin{array}{cc}
0 & \frac{\mu _{-}-\mu _{+}}{\lambda _{-}\lambda _{+}} \\ 
\frac{\mu _{+}-\mu _{-}}{\lambda _{-}\lambda _{+}} & 0%
\end{array}\right].
\end{eqnarray}
The matrix elements $L_{12}=L_{21}=\left({\mu _{-}+\mu _{+}}\right)/{\lambda_{-}\lambda _{+}}$ induce the Stark shift of the energies, which can be fully taken into account at the expense of lacking analytical expressions for the eigenenvalues and will be neglected in the following analytical derivation~\cite{Braak2011}. Moreover, we also neglect the counter-rorating-wave terms $\sigma_{+}b^{\dagger}+\sigma_{-}b$. Then the total Hamiltonian can be finally given in the generalized rotating-wave approximation (GRWA)~\cite{Irish2007}, as 
%\begin{widetext}
\begin{eqnarray}
H_{3}^{\rm{GRWA}} &=&\omega _{m}b^{\dagger }b+{\xi} _{-,{N}_{b}}
\left\vert \frac{1}{2},-\frac{1}{2}\right\rangle \left\langle \frac{1}{2},-%
\frac{1}{2}\right\vert \nonumber \\
&&+{\xi} _{+,{N}_{b}} \left\vert \frac{1}{2},\frac{1}{2}%
\right\rangle \left\langle \frac{1}{2},\frac{1}{2}\right\vert \nonumber \\
&&+{\mathcal{P}}_{{N}_{b}}b^{\dagger }\left\vert \frac{1}{2},-\frac{1}{2}\right\rangle
\left\langle \frac{1}{2},\frac{1}{2}\right\vert \nonumber \\
&&+{\mathcal{P}}_{{N}_{b}} b\left\vert \frac{1}{2},%
\frac{1}{2}\right\rangle \left\langle \frac{1}{2},-\frac{1}{2}\right\vert, \label{H_{3}^{GRWA}}
\end{eqnarray}
%\end{widetext}
with the phonon number dependent parameters
\begin{eqnarray}
{\xi} _{\pm,{N}_{b}}=\varepsilon _{\pm}+ 
\frac{\mu _{\pm}}{\lambda _{\pm}^{2}}\left({\omega _{A}-\omega _{B}}\right)\sin \phi \left[ G_{0}\left( b^{\dagger
}b\right) -\beta \right],
\end{eqnarray}
and
\begin{eqnarray}
{\mathcal{P}}_{{N}_{b}}=\frac{1}{2}M_{12}\left(\omega
_{A}-\omega _{B}\right) \sin \phi F_{1}\left( b^{\dagger }b\right).
\end{eqnarray}
Here, the superscript GRWA refers to the fact that the rotating-wave approximation is made after performing the first-order correction. The rotating-wave term for the expansion of $i\sigma_{y}\sinh\left[ G\left(b^{\dagger}-b \right)/\omega_m \right]$ in Eq.~(\ref{H3}) is exhibited in the energy-conserving terms $b\left\vert \frac{1}{2},%
\frac{1}{2}\right\rangle \left\langle \frac{1}{2},-\frac{1}{2}\right\vert+ \rm{h.c.}$ with phonon number dependent coupling strength $\mathcal{P}_{{N}_{b}}$. The energy-conserving terms also indicate the transition between different phonon number excitations. In the basis of $%
\left\vert \frac{1}{2},-\frac{1}{2},n_{b}\right\rangle $ and $\left\vert 
\frac{1}{2},\frac{1}{2},n_{b}-1\right\rangle $ $\left( n_{b}=1,2,\cdots
\right) ,H_{3}^{\rm{GRWA}}$ takes the following matrix form%
\[
H_{3,n_b}^{\rm{GRWA}}=\left[ 
\begin{array}{cc}
n_{b}\omega _{m}+\xi _{-,n_{b}} &\mathcal{P} \\ 
\mathcal{P} & \left(
n_{b}-1\right) \omega _{m}+\xi _{+,n_{b}-1}%
\end{array}%
\right], 
\]
with
\begin{eqnarray*}
\xi _{\pm,n_{b}}&=&\varepsilon _{\pm}+ \frac{\mu _{\pm}}{\lambda _{\pm}^{2}}\left({\omega _{A}-\omega _{B}}\right)\sin \phi \left[ G_{0}\left(
n_{b}\right) -\beta \right],\\
\mathcal{P}&=&\frac{1}{2}M_{12}\left( {\omega _{A}-\omega _{B}}\right) \sin \phi
R_{n_{b}-1,n_{b}},\\
 R_{n_{b}-1,n_{b}}&=&\left\langle n_{b}\right\vert F_{1}\left( b^{\dagger
}b\right) b^{\dagger}\left\vert n_{b}-1\right\rangle \nonumber\\
&=&\frac{1}{\sqrt{n_{b}}}\frac{G }{\omega _{m}}\exp\left({-%
\frac{G^2}{2\omega _{m}^{2}}}\right)L_{n_{b}-1}^{1}%
\left(\frac{G^2}{\omega _{m}^{2}}\right). 
\end{eqnarray*}
Thus, the eigenenergies of the system in the case of single phonon exchange can be given as 
%\begin{widetext}
%\begin{eqnarray*}
%E_{\frac{1}{2},p,n_{b}}^{GRWA}=\frac{1}{2}\left[ 
%\begin{array}{c}
%\left( 2n_{b}-1\right) \omega _{m}+\xi _{-,n_{b}}+\xi _{+,n_{b}-1} \pm \sqrt{\left( \omega _{m}+\xi _{-,n_{b}}-\xi _{+,n_{b}-1}\right)
%^{2}+4\left( \frac{\mu _{-}-\mu _{+}}{\lambda _{-}\lambda _{+}}\frac{\left(
%\omega _{A}-\omega _{B}\right) }{2}\sin \phi R_{n_{b}-1,n_{b}}\right) ^{2}}%
%\end{array}%
%\right]  \\
%\end{eqnarray*}
%\end{widetext}
\begin{eqnarray}
E_{\frac{1}{2},p,n_{b}}^{\rm{GRWA}}&=&
\left( n_{b}-\frac{1}{2}\right) \omega _{m}+\frac{1}{2}(\xi _{-,n_{b}}+\xi _{+,n_{b}-1} )\nonumber\\
&&\pm\frac{1}{2} \sqrt{\left( \omega _{m}+\xi _{-,n_{b}}-\xi _{+,n_{b}-1}\right)
^{2}+4{\mathcal{P}}^{2}}.  \label{EGRWA}
\end{eqnarray} 
Besides, the corresponding eigenfunctions are given by
\begin{eqnarray*}
\left\vert \varphi _{\frac{1}{2},p,n_{b}}^{\rm{GRWA}}\right\rangle =\frac{1}{%
t_{\frac{1}{2},p,n_{b}}}\left( \left\vert \frac{1}{2},-\frac{1}{2}%
,n_{b}\right\rangle +\nu _{\frac{1}{2},p,n_{b}}\left\vert \frac{1}{2},\frac{1%
}{2},n_{b}-1\right\rangle \right), 
\end{eqnarray*}
with 
\begin{eqnarray}
 \nu _{\frac{1}{2},n_{b},p} &=&{\mathcal{P}}^{-1}[E_{\frac{1}{2},p,n_{b}}^{\rm{GRWA}}-\left(
n_{b}\omega _{m}+\xi _{-,n_{b}}\right)],\\
t_{\frac{1}{2},p,n_{b}}&=&\sqrt{1+\nu _{%
\frac{1}{2},n_{b},p}^{2}}.
\end{eqnarray}
The ground-state energy for the state $\left\vert \frac{1}{2},-\frac{1}{2},0\right\rangle$ is $E_{G}^{\rm{GRWA}}=\varepsilon _{-}$. When $\lambda=g_0$, Eq.~(\ref{EGRWA}) can be simplified into  
\begin{equation}
E_{\frac{1}{2},p,n_{b}}^{\rm{GRWA}} =\left( n_{b}-\frac{1}{2}\right) \omega _{m}-\frac{g_{0}^{2}}{\omega _{m}}\pm\frac{1}{2}\left\vert \omega _{m}-\left\vert\frac{2\eta }{\sin 2\theta}\right\vert\right\vert.   \label{Eequal}
\end{equation}
As we know, unitary transformation has nothing to do with the eigenengergies but changes the eigenfunctions. Thus the eigenfunctions, corresponding to $\mathcal{N}=1$, for the original Hamiltonian $H_{1}$ are given by
\begin{eqnarray}
\left\vert \psi _{\frac{1}{2},p,n_{b}}^{\rm{GRWA}}\right\rangle&=&U_{1}^{\dagger
}U_{2}^{\dagger }U_{3}^{\dagger }\left\vert \varphi _{\frac{1}{2}%
,p,n_{b}}^{\rm{GRWA}}\right\rangle  \nonumber\\
&=&U_{1}^{\dagger }
\left\vert \frac{1}{2},-\frac{1}{2}\right\rangle \frac{1}{\lambda _{-}%
}\frac{1}{t_{\frac{1}{2},p,n_{b}}}\left\vert n_{b}\right\rangle _{\frac{1}{2},-\frac{1}{2%
}} \nonumber\\
&&+U_{1}^{\dagger }
\left\vert \frac{1}{2},-\frac{1}{2}\right\rangle \frac{1}{\lambda _{+}}\frac{\nu _{\frac{1}{2},n_{b},p}}{t_{\frac{1}{2}%
,p,n_{b}}}\left\vert n_{b}-1\right\rangle _{\frac{1}{2},-\frac{1}{2}} \nonumber\\
&&+U_{1}^{\dagger }\left\vert \frac{1}{2},\frac{1}{2}\right\rangle  \frac{\mu _{-}}{%
\lambda _{-}}\frac{1}{t_{\frac{1}{2},p,n_{b}}}\left\vert n_{b}\right\rangle
_{\frac{1}{2},\frac{1}{2}}\nonumber\\
&&+U_{1}^{\dagger }\left\vert \frac{1}{2},\frac{1}{2}\right\rangle\frac{\mu _{+}}{\lambda _{+}}\frac{\nu _{\frac{1}{2},n_{b},p}%
}{t_{\frac{1}{2},p,n_{b}}}\left\vert n_{b}-1\right\rangle _{\frac{1}{2},\frac{1}{2}}  \label{Wavefunction}\\
\left\vert \psi _{G}^{\rm{GRWA}}\right\rangle &=&U_{1}^{\dagger }U_{2}^{\dagger
}\left\vert \frac{1}{2},-\frac{1}{2},0\right\rangle =U_{1}^{\dagger
}\left\vert \frac{1}{2},-\frac{1}{2}\right\rangle \left\vert 0\right\rangle
_{\frac{1}{2},-\frac{1}{2}} \nonumber\\  \label{WavefunctionG}
\end{eqnarray}
Here $\left\vert n_{b}\right\rangle _{\frac{1}{2}, m}=\exp\left[{-{\left(
\Omega +mG\right) }\left( b^{\dagger }-b\right)}/\omega _{m}\right]\left\vert
n_{b}\right\rangle$ with $m=1/2$ or $m=-1/2$ denotes the displaced Fock states with $n_{b}=1,2,\cdots$, and the parameter $\left(\Omega+mG\right)/\omega_{m}$ denotes the displacement determined by the quantum number $m$ in the $\mathcal{N}=1$ subspace.

Till now, we have gotten all the eigenvalues for the $\mathcal{N}=1$ subspace with GRWA. This method can be extended to other subspaces with higher polariton excitation number $\mathcal{N}$. In the Appendix A, the eigenenergies and eigenfunctions $E_{1,q,n_{b}}^{\rm{GRWA}}$ in Eq.~(\ref{wavefunction1qnb}) and $\left\vert \psi _{1,q,n_{b}}^{\rm{GRWA}}\right\rangle$ in Eq.~(\ref{E1qnb}) for the $\mathcal{N}=2$ subspace are also obtained. 

In Fig.~\ref{Figure2}, the eigenenergies in the two-, one-, zero-polariton subspace are plotted as a function of the photon-phonon coupling strength $g_{0}$, respectively. In each panel, we have subtracted the base energy $j(\omega_A+\omega_B)$, with $j=\mathcal{N}/2$. The energy level structures described by Eq.~(\ref{EGRWA}) for the resonant case $\omega_c=\omega_{ex}$ are shown in red-dotted curves, while the energy structures with the numerically exact diagonalization of the Hamiltonian in Eq.~(\ref{H2}) for each polariton subspace are shown in black-solid curves. The coincidence between the theoretical method and numerical one shows the validity of GRWA in the regime we are working with. With the increase of the coupling strength $g_0$, small discrepancies occur. They are mainly caused by the overlooked Stark effect and higher order phonon transitions. It is obvious that the energy levels display much more abundant nonlinearity compared to the large photon-exciton detuning case as shown in Eq.~(\ref{Energy}), which is caused by the coupling between different phonon number manifolds. What is more, as shown in the vertical black-dotted line in Figs.~\ref{Figure2}(a) and (b), the specific photon-phonon coupling strength $g_0$, where the energy gap $E_{\frac{1}{2},+,n_{b}}^{\rm{GRWA}}-E_{\frac{1}{2},-,n_{b}}^{\rm{GRWA}}$ (Eq.~(\ref{EGRWA})) in the same $n_b$-th manifold has the minimum value, is extremely close to the exciton-phonon coupling strength $\lambda$. If we further assume $\left\vert{2\eta}/{\sin 2\theta}\right\vert=\omega _{m}$, the gap is closed in the theoretical method, and the coupling strength $g_0$ equals $\lambda$. These phenomena can be used to detect the exciton-phonon coupling strength in the semiconducting cavity.

\section{EMISSION SPECTRA OF THE POLARITONS} \label{Sec.III}
Let us now study the mechanical effect on the emission spectra of polaritons. There are many loss mechanisms involved in the dynamics of this hybrid system, including the mechanical damping rate $\gamma_m$, polariton emission rates $\kappa_A$ and $\kappa_B$. However, in this section we only consider the simplest situation when the decay rates of the mechanical resonator and polariton modes are completely neglected (i.e., we set $\gamma_m=\kappa_A=\kappa_B=0$). Or equivalently, we assume the time length  $t$ of the excitation in the cavity satisfies the condition $1/\gamma \ll t\ll 1/\kappa_{A,B}\ll 1/\gamma_m$, where $\gamma $ is the half-bandwidth of the spectrometer, it is also reasonable to neglect the three decay mechanisms. Thus the only broadening mechanism comes from the detecting spectrometer, and its physical spectrum can be given by~\cite{Glauber1963, Eberly1977}
\begin{align}
S\left( \omega \right) =&2\gamma\int_{0}^{t}dt_{1}\int_{0}^{t}dt_{2} \exp\left[-\left( \gamma -i\omega \right)\left( t-t_{2}\right)\right]\nonumber\\
&\times \exp\left[-\left( \gamma +i\omega \right) \left(
t-t_{1}\right) \right]G\left( t_{1},t_{2}\right) 
\end{align}
where $G(t_{1},t_{2})$ represents the dipole correlation function of the polaritons and is defined as 
\begin{equation}
G\left( t_{1},t_{2}\right)=\left\langle \psi\left(0\right) \right\vert B^{\dagger }\left(t_{2}\right) B\left( t_{1}\right) \left\vert \psi\left(0\right)\right\rangle
\end{equation}
with $\left\vert \psi\left(0\right)\right\rangle$ the initial state of the system. Here we take the lower level polariton mode $B$ as an example, which can also be applied to the case of the mode $A$. Taking into account that the transition between different energy levels satisfies the condition
\begin{equation}
\left\langle j^{\prime }m^{\prime }\right\vert B\left\vert jm\right\rangle
=\sqrt{j-m-1}\delta _{j^{\prime },j-\frac{1}{2}%
}\delta _{m^{\prime },m+\frac{1}{2}},
\end{equation}
we can conclude the selection rule $j^{\prime }=j-\frac{1}{2}$ and $m^{\prime }=m+\frac{1}{2}$. And it is evident that $j^{\prime}$ is only determined by the initial state $\left\vert \psi\left(0\right) \right\rangle$. We first consider the case that the transition occurs between $\mathcal{N}=1$ and $\mathcal{N}=0$ subspaces and the initial state of the mechanical resonator is in the displaced Fock state $\left\vert n_{0}\right\rangle_{\frac{1}{2},-\frac{1}{2}}$. Thus we make the assumption that the initial state is written as $\left\vert \psi\left(0\right)\right\rangle =\left\vert \frac{1}{2},-\frac{1}{2}\right\rangle\left\vert n_{0}\right\rangle_{\frac{1}{2},-\frac{1}{2}} $. In fact, the method we used here is not restricted to our assumption of the initial state, but can be extended to more general case. The time evolution operator $U(t)$ of the system concerning these subspaces can be gotten from the eigenenergies and eigenstates, i.e. Eq.~(\ref{EGRWA}), Eq.~(\ref{Wavefunction}), and Eq.~(\ref{WavefunctionG}), which we have shown in the last section, that is,
\begin{eqnarray}
U\left( t\right)&=&e^{-iH_{1}t}=e^{-iE_{G}^{\rm{GRWA}}t}\left\vert \psi _{G}^{\rm{GRWA}}\right\rangle \left\langle   \psi _{G}^{\rm{GRWA}}   \right\vert  \nonumber\\
&&+\sum\limits_{p,n_{b}}e^{-iE_{\frac{1}{2},p,n_{b}}^{\rm{GRWA}}t} \left\vert \psi _{\frac{1}{2},p,n_{b}}^{\rm{GRWA}}\right\rangle \left\langle  \psi _{\frac{1}{2},p,n_{b}}^{\rm{GRWA}}\right\vert .\nonumber\\
\end{eqnarray}
Taking into account the fact that $B(t)=U^{\dagger}(t)BU(t)$, the correlation $G(t_{1},t_{2})$ can be obtained as
%\begin{widetext}
\begin{eqnarray*}
G\left( t_{1},t_{2}\right) &=&\sum\limits_{n_{1}=0,n_{2}}e^{i\left( E_{G}^{\rm{GRWA}}-E^{\rm{GRWA}}_{00n_{2}}\right) \left(
t_{2}-t_{1}\right) }\left\vert \left\langle i|\psi _{G}^{\rm{GRWA}}\right\rangle
\right\vert ^{2}\nonumber\\
&&\times
\left\vert \left\langle \psi _{G}^{\rm{GRWA}}\right\vert B^{\dagger
}\left\vert \psi _{00n_{2}}\right\rangle \right\vert ^{2} \nonumber\\
&&+\sum\limits_{n_{1}>0,p,n_{2}}e^{i\left( E^{\rm{GRWA}}_{\frac{1}{2}%
,p_{1},n_{1}}-E^{\rm{GRWA}}_{00n_{2}}\right) \left( t_{2}-t_{1}\right) }\nonumber\\
&&\times\left\vert
\left\langle i|\psi ^{\rm{GRWA}}_{\frac{1}{2},p_{1},n_{1}}\right\rangle \right\vert
^{2}\left\vert \left\langle \psi^{\rm{GRWA}} _{\frac{1}{2},p_{1},n_{1}}\right\vert
B^{\dagger }\left\vert \psi _{00n_{2}}\right\rangle \right\vert ^{2},
\end{eqnarray*}%
%\end{widetext}
with $\left\vert \psi _{00n_{2}}\right\rangle=\left\vert 00\right\rangle\left\vert n_{2}\right\rangle$. Thus the stationary spectrum can be decomposed into three parts as
 \begin{equation}
 S_{B}^{10}\left( \omega \right)=S_{1}\left( \omega \right)+S_{2}\left( \omega \right)+S_{3}\left( \omega \right),
 \end{equation} 
 where
 \begin{eqnarray*}
 S_{1}\left( \omega \right)&=&\sum\limits_{n_{1}=0,n_{2}}\Gamma_{1}(\omega)\left\vert \left\langle \psi\left(0\right) |\psi _{G}^{\rm{GRWA}}\right\rangle \right\vert
^{2}\nonumber\\
&&\times\left\vert \left\langle \psi _{G}^{\rm{GRWA}}\right\vert B^{\dagger }\left\vert \psi
_{00n_{2}}\right\rangle \right\vert ^{2}, \\
S_{2}\left( \omega \right)&=&\sum\limits_{n_{1}>0,n_{2}}\Gamma_{2}(\omega)\left\vert \left\langle \psi\left(0\right) |\psi
^{\rm{GRWA}}_{\frac{1}{2},+,n_{1}}\right\rangle \right\vert ^{2}\\
&&\times
\left\vert
\left\langle \psi ^{\rm{GRWA}}_{\frac{1}{2},+,n_{1}}\right\vert B^{\dagger
}\left\vert \psi _{00n_{2}}\right\rangle \right\vert ^{2},\\
S_{3}\left( \omega \right)&=&\sum\limits_{n_{1}>0,n_{2}}\Gamma_{3}(\omega)\left\vert \left\langle \psi\left(0\right) |\psi
^{\rm{GRWA}}_{\frac{1}{2},-,n_{1}}\right\rangle \right\vert ^{2}\\
&&\times
\left\vert
\left\langle \psi ^{\rm{GRWA}}_{\frac{1}{2},-,n_{1}}\right\vert B^{\dagger
}\left\vert \psi _{00n_{2}}\right\rangle \right\vert ^{2}, 
\end{eqnarray*}
\begin{figure}[ptb]
\includegraphics[scale=0.4,clip]{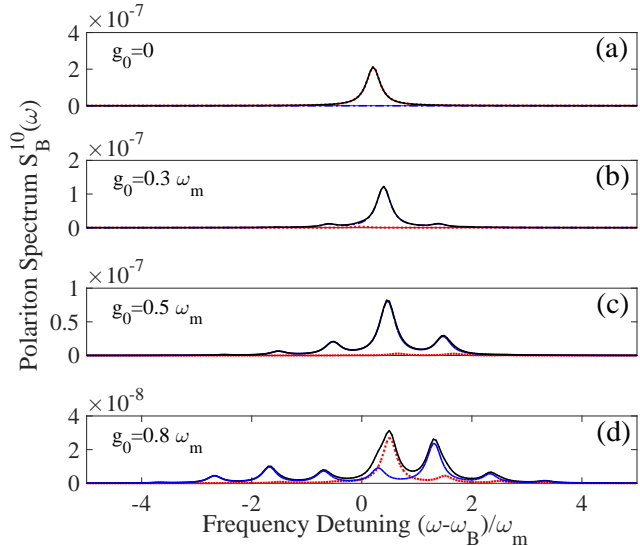}\
%bb=20 220 550 610,  width=0.45\textwidth, 
%\includegraphics[scale=0.25,clip]{fig1b.jpeg}%bb=25 220 573 640, scale=0.5,
 \caption{(Color online) Emission spectrum of polariton mode B (black solid curves) from $\mathcal{N}=1$ to $\mathcal{N}=0$ as a function of scaled frequency detuning $\left(\omega-\omega_{B}\right)/\omega_m$ without the exciton-phonon coupling, i.e., $\lambda=0$, under different cavity photon-phonon coupling strength $g_0=0$, 0.3 $\omega_m$, 0.5 $\omega_m$, 0.8 $\omega_m$ from (a) to (d). The black solid, blue dash-dotted and red dotted curves represent the $S_B^{10}(\omega)$, $S_2(\omega)$ and $S_3(\omega)$, respectively. Other parameters are set to be: $\Delta_{ce}=0$, $\eta$=0.5 $\omega_m$, $\gamma$=0.15 $\omega_m$.}%
\label{Figure3}%
\end{figure}
with 
\begin{equation}
\Gamma_{i}(\omega)=\frac{2\gamma }{\gamma ^{2}+\left[ \omega
-\omega _{B}-\left(\delta_i-n_{2}\omega_m\right) %
\right] ^{2}},
\end{equation}
and $\delta_1={\eta}/{\sin 2\theta }+E_G^{\rm{GRWA}}$, $\delta_2={\eta}/{\sin 2\theta }+E^{\rm{GRWA}}_{\frac{1}{2},+,n_{1}}$, $\delta_3={\eta}/{\sin 2\theta }+E^{\rm{GRWA}}_{\frac{1}{2},-,n_{1}}$. Physically, this decomposition can be understood by the fact that the initial state $\left\vert \psi\left(0\right)\right\rangle =\left\vert \frac{1}{2},-\frac{1}{2}\right\rangle\left\vert n_{0}\right\rangle_{\frac{1}{2},-\frac{1}{2}}$ is not the eigenstate of the Hamiltonian in Eq.~(\ref{H1}), but it can always be interpreted as the superposition of the eigenstates $\left\vert \psi _{G}^{\rm{GRWA}}\right\rangle$ and $\left\vert \psi ^{\rm{GRWA}}_{\frac{1}{2},p,n_1}\right\rangle$ ($p=\pm$). Moreover, the subscript in $S_{B}^{10}\left( \omega \right)$ denotes that this is the emission spectrum for the lower level polariton mode $B$, while the superscript denotes the transition is from $\mathcal{N}=1$ subspace to $\mathcal{N}=0$ subspace. Note the transient terms and very slowly variation terms have been neglected and the base line $\left(\omega _{A}+\omega _{B}\right)/{2}$ for the $\mathcal{N}=1$ subspace is added. We can observe that the eigenvalues determine the positions of the spectral component and the ovelap between different states decides the intensity of the spectral lines. As a matter of fact, the spectrum is composed of three parts with equidistance but different central points $\delta_1$, $\delta_2$, $\delta_3$. This is different from the results shown in Refs.~\cite{Rabl2011, Liao2012}. For each part, the interval is marked by the mechanical resonator frequency $\omega_{m}$, and $(n_{1}-n_{2})$ with $n_{1}, n_{2}\in\left[0,\infty\right)$ gives us a clue for numerous sidebands. These sidebands are developed around $\delta_{1}$, $\delta_{2}$, $\delta_3$ respectively, and semantically we name them as center frequencies. However, the sidebands can only be resolved when their peaks go over the height of nearby Lorentzian. 

\begin{figure}[ptb]
\includegraphics[scale=0.4,clip]{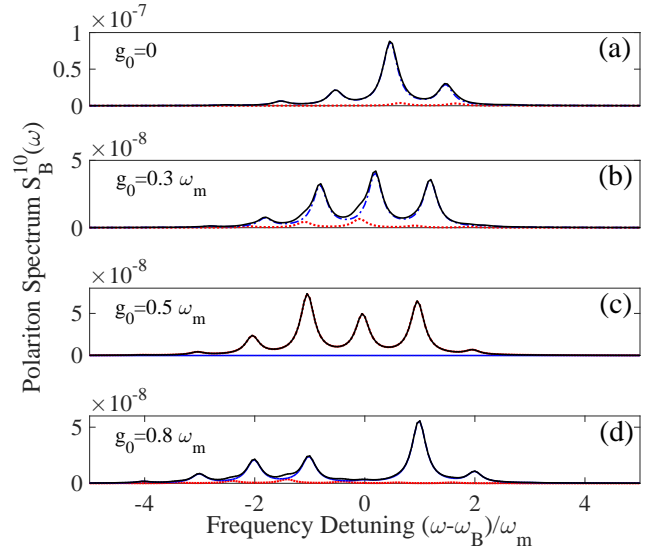}\
%bb=20 220 550 610,  width=0.45\textwidth, 
%\includegraphics[scale=0.25,clip]{fig1b.jpeg}%bb=25 220 573 640, scale=0.5,
 \caption{(Color online) Emission spectrum of polariton mode B (black solid curves) from $\mathcal{N}=1$ to $\mathcal{N}=0$ as a function of scaled frequency detuning $\left(\omega-\omega_{B}\right)/\omega_m$ with the exciton-phonon coupling $\lambda$=0.5 $\omega_m$, under different cavity photon-phonon coupling strength $g_0=0$, 0.3 $\omega_m$, 0.5 $\omega_m$, 0.8 $\omega_m$ from (a) to (d). The black solid, blue dash-dotted  and red dotted curves represent the $S_B^{10}(\omega)$, $S_2(\omega)$ and $S_3(\omega)$, respectively. Other parameters are set to be: $\Delta_{ce}=0$, $\eta$=0.5 $\omega_m$, $\gamma$=0.15 $\omega_m$.}%
\label{Figure4}%
\end{figure}

In Fig.~\ref{Figure3}, the emission spectrum $S_{B}^{10}(\omega)$ (black-solid curves) is plotted as a function of the frequency detuning $(\omega-\omega_B)/\omega_m$ with various optomechanical coupling strengths $g_{0}=0$, 0.3 $\omega_m$, 0.5 $\omega_m$, 0.8 $\omega_m$. In order to see the effect of the mechanical resonator clearly, we first exclude the influence of exciton-phonon coupling by setting $\lambda=0$. Spectra $S_{2}(\omega)$ and $S_{3}(\omega)$ with central frequencies $\delta_2$, $\delta_3$ are plotted in the blue-dash-dotted and the red-dotted curves, respectively. The $S_{1}(\omega)$ does not show up here because we choose $n_0=2$. In this case, the initial state $\left\vert \psi\left(0\right)\right\rangle =\left\vert \frac{1}{2},-\frac{1}{2}\right\rangle\left\vert 2\right\rangle_{\frac{1}{2},-\frac{1}{2}} $ is orthogonal to the ground state $\left\vert\psi _{G}^{\rm{GRWA}}\right\rangle$, i.e, $\left\langle \psi\left(0\right) |\psi _{G}^{\rm{GRWA}}\right\rangle=0$, which leads to $S_{1}(\omega)=0$. First, in Fig.~\ref{Figure3}(a), we consider the situation $g_0=0$, i.e., the polartion mode is totally decoupled with the mechanical resonator. Only one Lorentzian peak appears around $\omega=\omega_B$ denoting the bare polariton mode spectrum, which is not affected by the mechanical resonator. And it mainly comes from the contribution of $S_{3}(\omega)$. Moreover, with the increase of optomechanical coupling strength $g_0$ from 0.3 $\omega_m$ to 0.8 $\omega_m$, as shown in Fig.~\ref{Figure3}(b) to Fig.~\ref{Figure3}(d), we find that the contributions of $S_{2}(\omega)$ and $S_{3}(\omega)$ to the total spectrum $S_{B}^{10}(\omega)$ vary a lot. Besides, more sidebands appear at the frequency $\delta_2-n_2\omega_m$ and $\delta_3-n_2\omega_m$ for $g_0>\gamma_c$, e.g., two to six sidebands from Fig.~\ref{Figure3}(b) to Fig.~\ref{Figure3}(d), while at the expense of lower central peak. Usually, the maximum number of sidebands corresponds to the phonon number truncated for calculation (here we set as 6). 

When the exciton-phonon coupling is included, as shown in Fig.~\ref{Figure4} with the strength $\lambda$=0.5 $\omega_m$. For each specific $g_0$, more sidebands appear, compared to the case of $\lambda=0$ as shown in Fig.~\ref{Figure3}. Even when $g_0=0$, there are three sidebands as shown in Fig.~\ref{Figure4}(a) and it mainly comes from the contribution of $S_2(\omega)$. From Fig.~\ref{Figure3} and Fig.~\ref{Figure4}, we can find that the mechanical resonator enriches the spectrum of polariton mode with more sidebands through the coupling with both the exciton and cavity photon modes.

\section{BUNCHING AND ANTI-BUNCHING RESONANCES OF THE POLARITON MODES}\label{Sec.IV}
In this section, we will show how the mechanical resonator affects the statistical properties of the polariton emission. We assume that the cavity field of the hybrid system is driven by a weak classical field with the frequency $\omega_d$. In this case, the Hamiltonian in Eq.~(\ref{H}) is changed into 
\begin{equation}
H^{'}=H+i\varepsilon\left(a^{\dagger}e^{-i\omega_d t}-a e^{i\omega_d t} \right).
\end{equation}
Here $H$ is given in Eq.~(\ref{H}), and $\varepsilon$ is the coupling strength between the cavity field and the driving field. In the rotating reference frame under the frequency $\omega_d$ of the driving field with unitary operator $R(t)=\exp[-i\omega_d(A^{\dagger}A+B^{\dagger}B)t]$ of the polariton operators, we can write the total Hamiltonian of the system as
\begin{eqnarray}
 \widetilde{H}^{\prime} &=&\Delta _{A}A^{\dagger }A+\Delta _{B}B^{\dagger}B
+\omega _{m}b^{\dagger }b \nonumber\\
&&+ \left (Q_{A}A^{\dagger }A+Q_{B}B^{\dagger }B \right)\left( b^{\dagger }+b\right)\nonumber\\
&&+ Q\left( A^{\dagger }B +B^{\dagger }A\right)\left( b^{\dagger }+b\right)\nonumber\\
&&+i\varepsilon \left[ \cos \theta \left( A^{\dagger }-A\right)   -\sin \theta \left(B^{\dagger }-B\right)\right] \label{Heff} 
\end{eqnarray}
where $\Delta_A=\omega_A-\omega_d$ $(\Delta_B=\omega_B-\omega_d)$ is the detuning between the polariton mode $A$ ($B$) and the driving field. For the open system, the dissipative terms in the polariton representation must be considered and can be expressed in the Lindblad superoperator form
\begin{eqnarray}
 \mathcal{L}_{diss}&\simeq&\frac{\gamma_{m}}{2} \left[\left(N_{\rm{th}}+1\right)\mathcal{D}\left[b\right]+N_{\rm{th}}\mathcal{D}\left[b^{\dagger}\right] \right] \nonumber\\
  &&+\kappa_{A}\mathcal{D}\left[A\right]+\kappa_{B}\mathcal{D}\left[B\right]. \label{Ldiss}
\end{eqnarray}
Here, the superoperator has the form of $\mathcal{D}\left[o\right]\rho=o\rho o^{\dagger}-\frac{1}{2}\left(o^{\dagger}o\rho+\rho o^{\dagger}o\right)$ ($o$ can be any operator of the system, i.e., $A, B, b$).  The first line in the Eq.~(\ref{Ldiss}) describes the coupling of the mechanical resonator to a thermal bath, and $N_{\rm{th}}=1/[\exp(\hbar\omega_{m}/k_{B}T)-1]$ denotes the thermal phonon number at temperature $T$, with $k_{B}$ the Boltzmann constant. $\mathcal{D}\left[A\right], \mathcal{D}\left[B\right]$ represent the leakage of the polariton modes $A$ and $B$ with the polariton decay rates 
\begin{eqnarray}
\kappa_{A}=\kappa_{a}\cos^{2}\theta +\kappa_{ex}\sin^{2}\theta,\label{KA} \\ 
\kappa_{B}=\kappa_{a}\sin^{2}\theta +\kappa_{ex}\cos^{2}\theta. \label{KB}
\end{eqnarray}
\begin{figure}[ptb]
\includegraphics[scale=0.4,clip]{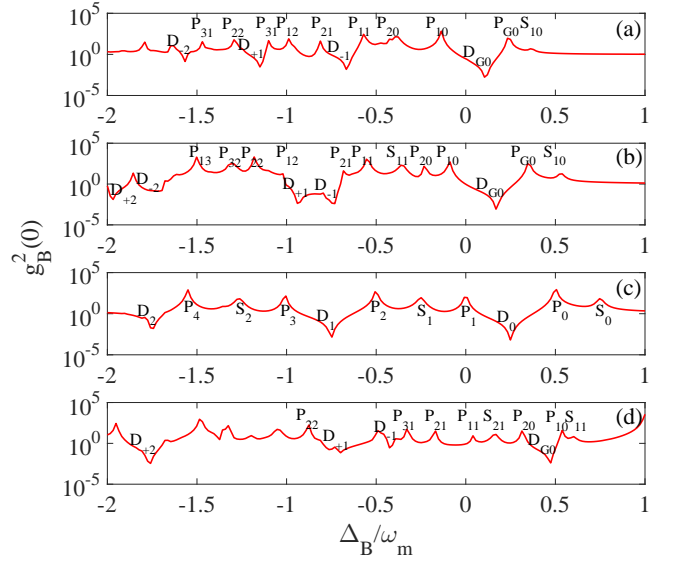}\
%bb=20 220 550 610,  width=0.45\textwidth, 
%\includegraphics[scale=0.25,clip]{fig1b.jpeg}%bb=25 220 573 640, scale=0.5,
 \caption{(Color online) Dependence of the equal-time second-order correlation $g^{(2)}(0)$ on the rescaled detuning $\Delta_B/\omega_m$ for various coupling strengths, e.g., $g_0=0$ in (a), $g_0$=0.3 $\omega_m$ in (b), $g_0$=0.5 $\omega_m$ in (c), and $g_0$=0.8 $\omega_m$ in (d). Other parameters are $\Delta_{ce}=0$, $\lambda$=0.5 $\omega_m$, $\eta$=0.5 $\omega_m$. }%
\label{Figure5}%
\end{figure}
Here, $\kappa_a$, $\kappa_{ex}$ and $\gamma_m$ represent the decay rates of the cavity, exciton and mechanical modes respectively. Besides, we note that the term $(\kappa_a-\kappa_{ex})^2$ has been neglected in the process of getting the decay rates $\kappa_A$ and $\kappa_B$ as shown in Eq.~(\ref{KA}) and Eq.~(\ref{KB}). This is reasonable when the splittings of these two modes are far larger than their decay rates $\kappa_{a}, \kappa_{ex}$, which is consistent with our original assumption. Because of the high frequency of the polariton modes, we have neglected the thermal excitations of excitons and photons in the low-temperature limit. Then the master equation for the reduced density matrix operator $\rho$ of the whole system can be described by 
 \begin{equation}
\dot{\rho}=i\left[ \rho,\widetilde{H}^{\prime} \right]+\mathcal{L}_{diss}\rho. 
\end{equation}
It can be solved numerically in the complete basis set $\left\vert n_{A}\right\rangle\otimes \left\vert n_{B}\right\rangle\otimes \left\vert n_{b}\right\rangle$, with $n_{A}, n_{B}$ and $n_{b}=0,1,2\cdots$ denoting the excitation number in polariton $A, B$ and  mechanical resonator modes, respectively. In this work, the numerical calculations by solving the master equation within a truncated Fock state space are done by using the quantum toolbox~\cite{Johansson2012,Johansson2013}. 

Next we use polariton mode B as an example to show the effect of the mechanical resonator on the statistical properties of the polaritons. The equal-time second-order correlation function of polariton mode $B$ can be given as~\cite{Scully1997}
\begin{equation}
g^{(2)}(0)=\frac{\left\langle B^{\dagger}B^{\dagger}BB\right\rangle}{ \left\langle B^{\dagger}B\right\rangle^2}, \label{g2BB}
\end{equation}
which describes the statistical properties of the polariton mode B. The status $g^{(2)}(0)<1$ ($g^{(2)}(0)>1$) characterizes the poloriton blockade (tunneling) process,~\cite{Tian1992,Imamoglu1997,Birnbaum2005,Miranowicz2013,Xunwei2013,Xu2013,Liu2014,Liu2016} in which the polariton exhibits sub-Poisson (or super-Poisson) statistics.  

Figure~\ref{Figure5} shows how the polariton statistics depends on dimensionless detuning $\Delta_B/\omega_m$. We note that each panel has several dips and resonant peaks, which denote the one-polariton and multi-polariton resonant transition, respectively. 

It will be easier to understand from the coupling balanced case $g_0=\lambda$, which is shown in Fig.~\ref{Figure5}(c). We only need one subscript $n_b$ ($n_b=0,1,2\cdots$) to label the dips and peaks caused by different phonon number. Specifically, the dips $D_{n_b} (n_b=0,1,2\cdots)$ are caused by the single polariton transition from the state $\left\vert \frac{1}{2},-\frac{1}{2}\right\rangle \left\vert{n}_{b}\right\rangle_{\frac{1}{2},-\frac{1}{2}}$ to the state $\left\vert0,0\right\rangle\left\vert 0\right\rangle$, when the detuning between the driving field and the polariton mode B satisfies the condition $\Delta_B/\omega_m=(g_0^2/\omega_m^2)-n_b$. The peaks $P_{n_b} (n_b=0,1,2\cdots)$ correspond to the two-polariton transition from the state $\left\vert 1, -1\right\rangle \left\vert{n}_{b}\right\rangle_{1,-1}$ to the state $\left\vert0,0\right\rangle\left\vert0\right\rangle$ when $\Delta_B/\omega_m=(2g_0^2/\omega_m^2)-(n_b/2)$. Moreover, we also label another series of peaks $S_{n_b} (n_b=0,1,2\cdots)$, which is the consequence of one-polarion transition from $\left\vert 1, -1\right\rangle \left\vert{n}_{b}\right\rangle_{1,-1}$ to $\left\vert\frac{1}{2},-\frac{1}{2}\right\rangle\left\vert0\right\rangle_{\frac{1}{2},-\frac{1}{2}}$ on the condition of $\Delta_B/\omega_m=(3g_0^2/\omega_m^2)-n_b$. Obviously, the distribution of these points is equally spaced, and all separated by one time or half of the frequency of the mechanical resonator $\omega_m$ in each series. 

When $g_0\ne\lambda$, the exchange interaction between $A, B$ and $b$ makes the system exhibit richer nonlinearity, and the wave eigenfunction concerning the phonon changes from the displaced Fock-state $\left\vert{n}_{b}\right\rangle_{j,m}$ to their superposition as shown in Eq.~(\ref{Wavefunction}). Thus the positions of the dips and peaks change a lot and we introduce two subscripts to label them. As shown in Fig.~\ref{Figure5}(a), (b) and (d), we consider three different coupling strength $g_0=0$, 0.3 $\omega_m$ and 0.8 $\omega_m$, respectively, and figure out that:

(i) the dips labeled by $D_{p,n_b}$ in Figs.~\ref{Figure5}(a), (b) and (d) are the results of one-polariton transition from the state $\left\vert \psi _{\frac{1}{2},p,n_{b}}^{\rm{GRWA}}\right\rangle$ to the state $\left\vert0,0\right\rangle\left\vert 0\right\rangle$ at the effective detuning $\Delta_B^{\prime}=\Delta_B+\eta/\sin2\theta=-E_{\frac{1}{2},p,n_{b}}^{\rm{GRWA}}$. Note that the term $\eta/\sin2\theta$ is added, it is because we take the base line $(\omega_A+\omega_B)/2$ for the $\mathcal{N}=1$ subspace into account. The parameter $p=\pm$ when $n_b=1,2,3\cdots$, while $p=G$ when $n_b=0$. For example, the symbol $D_{G0}$ labels the transition from the ground state $\left\vert \psi_G^{\rm{GRWA}}\right\rangle$ in the $\mathcal{N}=1$ subspace to the state $\left\vert0,0\right\rangle\left\vert 0\right\rangle$. The equal-time second-order correlation functions $g^{(2)}(0)$ corresponding to the dips $D_{p,n_b}$ are smaller than 1, i.e., $g^{(2)}(0)<1$, which means that the probability to excite two-polariton is smaller than that to excite two single-polariton independently, and then the polariton blockade happens and exhibits sub-Poisson statistics.

\renewcommand\arraystretch{2.4}
\begin{table}[htp]
\centering
\caption{Symbols representing the transitions from the original states (the second column) to the final states (the third column) for the balanced coupling case $g_0=\lambda$. And the last column shows the corresponding conditions which the frequency detuning $\Delta_B$ satisfies, respectively. The symbol $D$ represents the series of dips, while $P, S$ represent two series of peaks with one subscript $n_b$.}
\begin{tabular}{|p{1.2cm}<{\centering}|p{2.3cm}<{\centering}|p{2.2cm}<{\centering}|p{1.8cm}<{\centering}|}
\hline
\thead{Symbol}  & \thead{Original state} & \thead{Final state} & \thead{ Detuning $\Delta_B$}  \\ 
\hline
$D_{n_b}$  &  $\left\vert \frac{1}{2},-\frac{1}{2}\right\rangle \left\vert{n}_{b}\right\rangle_{\frac{1}{2},-\frac{1}{2}}$ &$\left\vert0,0\right\rangle\left\vert 0\right\rangle$ & $\frac{g_0^2}{\omega_m^2}-n_b$  
\\
\hline
$P_{n_b}$   &  $\left\vert 1,-1\right\rangle \left\vert{n}_{b}\right\rangle_{1,-1}$ & $\left\vert0,0\right\rangle\left\vert 0\right\rangle$ &$\frac{2g_0^2}{\omega_m^2}-\frac{n_b}{2}$  \\
\hline
$S_{n_b}$   &  $\left\vert 1,-1\right\rangle \left\vert{n}_{b}\right\rangle_{1,-1}$ &  $\left\vert \frac{1}{2},-\frac{1}{2}\right\rangle \left\vert 0\right\rangle_{\frac{1}{2},-\frac{1}{2}}$ & $\frac{3g_0^2}{\omega_m^2}-n_b$ \\
\hline
\end{tabular}  \label{Table1}
\caption{Symbols representing the transitions from the original states (the second column) to the final states (the third column) for the unbalanced coupling case $g_0\ne\lambda$. And the last column shows the corresponding conditions which the effective frequency detuning $\Delta_B^{\prime}=\Delta_B+\eta/\sin2\theta$ satisfies, respectively. The symbol $D$ represents the series of dips, while $P, S$ represent two series of peaks, with two subscripts $p\ (q), n_b$.}
\begin{tabular}{|p{1.3cm}<{\centering}|p{1.9cm}<{\centering}|p{2.2cm}<{\centering}|p{2cm}<{\centering}|}
\hline
\thead{Symbol }  & \thead{Original state} & \thead{Final state} & \thead{Effective  \\  Detuning $\Delta_B^{\prime}$}  \\ 
\hline
$D_{p,n_b}$  & $\left\vert \psi _{\frac{1}{2},p,n_{b}}^{\rm{GRWA}}\right\rangle$ &$\left\vert0,0\right\rangle\left\vert 0\right\rangle$ & $-E^{\rm{GRWA}}_{\frac{1}{2},p,n_{b}}$  
\\
\hline
$P_{q,n_b}$   & $\left\vert \psi _{1,q,n_{b}}^{\rm{GRWA}}\right\rangle$ & $\left\vert0,0\right\rangle\left\vert 0\right\rangle$ &$-E^{\rm{GRWA}}_{1,q,n_{b}}/2$  \\
\hline
$S_{q,n_b}$   & $\left\vert \psi _{1,q,n_{b}}^{\rm{GRWA}}\right\rangle$ & $\left\vert\frac{1}{2},-\frac{1}{2}\right\rangle\left\vert0\right\rangle_{\frac{1}{2},-\frac{1}{2}}$ & $-E^{\rm{GRWA}}_{1,q,n_b}+E^{\rm{GRWA}}_{\frac{1}{2},-\frac{1}{2},0}$ \\
\hline
\end{tabular} \label{Table2}
%\caption{Symbols representing the transitions from the original states (the second column) to the final states (the third column). And the last column shows the corresponding conditions which the effective detuning $\Delta_B^{\prime}=\Delta_B+\eta/\sin2\theta$ satisfies, respectively. The symbol $D$ represents the series of dips, while $P, S$ represent two series of peaks.  For the unbalanced coupling case ($g_0\ne\lambda$), $p=\pm$ and $q=1,2,3$ when $n_b=1,2,3...$, while $p=G$ and $q=1,2$ when $n_b=0$. For the balanced coupling case ($g_0=\lambda$), we only need to use the subscript $n_b$.} \label{Table}
\end{table}%

(ii) The peaks marked by $P_{q,n_b}$ represent the transition from the state $\left\vert {\psi} _{1,q,n_{b}}^{\rm{GRWA}}\right\rangle$ in the subspace $\mathcal{N}=2$ to the state $\left\vert0,0\right\rangle\left\vert 0\right\rangle$ at the effective detuning $\Delta_B^{\prime}=-E^{\rm{GRWA}}_{1,q,n_{b}}/2$. Here $q=1,2,3$ when  $n_b=1,2,3\cdots$ denoting the three energy levels in the $n_b$-th block, while for the $n_b=0$ block, $q=1, 2$ and $G$, standing for the first-, second-excited state, and the ground state respectively. And note that we have taken the base line $\omega_A+\omega_B$ for the $\mathcal{N}=2$ subspace into account. Specifically, $P_{10}$ represents two-polariton transition from the first excited state $\left\vert \psi _{1,1,0}^{\rm{GRWA}}\right\rangle$ to $\left\vert0,0\right\rangle\left\vert 0\right\rangle$ at the effective detuning $-E^{\rm{GRWA}}_{1,1,0}/2$, while $P_{G0}$ represents two-polariton transition from the ground state $\left\vert \tilde{\psi} _{G}^{\rm{GRWA}}\right\rangle$ (Eq.~(\ref{GroundN2})) in the $\mathcal{N}=2$ subspace to $\left\vert0,0\right\rangle\left\vert 0\right\rangle$ at the effective detuning $-\tilde{E}^{\rm{GRWA}}_{G}/2$ (Eq.~(\ref{EGN2})). Correspondingly, the equal-time second-order correlation functions $g^{(2)}(0)$ at these peaks $P_{q,n_b}$ are larger than 1, i.e., $g^{(2)}(0)>1$, which means that the probability to excite two-polariton is larger than that to excite two single-polariton independently, and then the polariton tunneling happens and exhibits super-Poisson statistics.

(iii) Besides, the small peaks pointed out by $S_{q,n_b}$ show the polariton transition between the state $\left\vert \psi _{1,q,n_{b}}^{\rm{GRWA}}\right\rangle$ in the $\mathcal{N}=2$ subspace and the ground state $\left\vert\frac{1}{2},-\frac{1}{2}\right\rangle\left\vert0\right\rangle_{\frac{1}{2},-\frac{1}{2}}$ in $\mathcal{N}=1$ subspace. And the transition frequency satisfies the condition $\Delta_B^{\prime}=-E^{\rm{GRWA}}_{1,q,n_b}+E^{\rm{GRWA}}_{\frac{1}{2},-\frac{1}{2},0}$. For example, $S_{11}$($S_{10}$) denotes the transition from the state $\left\vert \psi _{1,1,1}^{\rm{GRWA}}\right\rangle$ ($\left\vert \psi _{1,1,0}^{\rm{GRWA}}\right\rangle$) in the $\mathcal{N}=2$ subspace to the ground state $\left\vert \psi _{\frac{1}{2},-\frac{1}{2},0}^{\rm{GRWA}}\right\rangle$ in the $\mathcal{N}=1$ subspace at the effective detuning $\Delta_B^{\prime}=-E^{\rm{GRWA}}_{1,1,1}\ (E^{\rm{GRWA}}_{1,1,0})+E^{\rm{GRWA}}_{\frac{1}{2},-\frac{1}{2},0}$.

\begin{figure}[ptb]
\includegraphics[scale=0.39,clip]{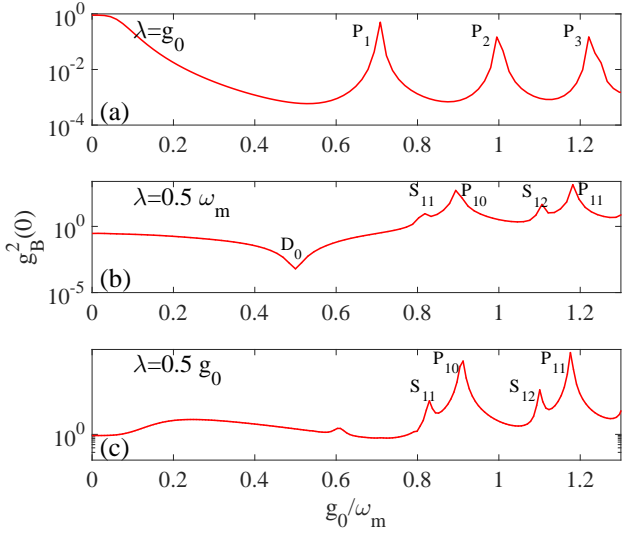}\
%bb=20 220 550 610,  width=0.45\textwidth, 
%\includegraphics[scale=0.25,clip]{fig1b.jpeg}%bb=25 220 573 640, scale=0.5,
 \caption{(Color online) Dependence of the equal-time second-order correlation $g^{(2)}(0)$ on the rescaled coupling strength $g_0/\omega_m$ with the pump detuning $\Delta_{B}=g_0^2/\omega_m$ for (a) coupling balanced $\lambda=g_0$ and detuning case: (b) $\lambda=0.5$ $\omega_m$, (c) $\lambda=0.5$ $g_0$. Other parameters are $\Delta_{ce}=0, \eta=0.5$ $\omega_m$. }%
\label{Figure6}%
\end{figure}

As a summary, in the first column of the TABLE~\ref{Table1} and TABLE~\ref{Table2}, we list the symbols which are used to label the transitions for the balanced ($g_0=\lambda$) and unbalanced ($g_0\ne\lambda$) coupling case, respectively. The transitions occur from the original states (the second column) to the final states (the third column), when the frequency detuning satisfies the conditions which are shown in the last column. We use two subscripts ($q,n_b$) for the unbalanced coupling case ($g_0\ne\lambda$) while only one ($n_b$) for the special balanced coupling case ($g_0=\lambda$) to discriminate different processes.

Furthermore, we show the equal-time second-order correlation $g^{(2)}(0)$ versus the rescaled radiation-pressure coupling strength $g_0/\omega_m$ in Fig.~\ref{Figure6} at the given detuning $\Delta_B=g_0^2/\omega_m$. As analyzed above, for the balanced coupling case $g_0=\lambda$, the single polariton transition from the state $\left\vert \frac{1}{2},-\frac{1}{2}\right\rangle \left\vert 0\right\rangle_{\frac{1}{2},-\frac{1}{2}}$ in the $\mathcal{N}=1$ subspace to the state $\left\vert0,0\right\rangle\left\vert 0\right\rangle$ occurs. And the polariton exhibits sub-Poisson statistics, i.e., the polariton blockade happens. However, the blockade is destroyed when the coupling strength satisfies the condition $g_0/\omega_m=\sqrt{n_b/2}$, which corresponds to the two-polariton transition from the state $\left\vert 1, -1\right\rangle \left\vert{n}_{b}\right\rangle_{1,-1}$ in the $\mathcal{N}=2$ subspace to the state $\left\vert0,0\right\rangle\left\vert0\right\rangle$, and can be seen in the peaks $P_{n_b}$ as shown in Fig.~\ref{Figure6}(a). For a specific exciton-phonon coupling strength, e.g., $\lambda$=0.5 $\omega_m$ in Fig.~\ref{Figure6}(b), polariton blockade occurs once $g_0=\lambda$, which can be seen in the dip labeled by $D_{0}$. With the increase of $g_0$, other two types of resonant transition occurs, as labeled by $P_{p,n_b}$  and $S_{q,n_b}$. Moreover, for the case of $\lambda$=0.5 $g_0$, as depicted in Fig.~\ref{Figure6}(c), we observe resonant peaks labeled by $S_{11}$, $S_{21}$, $P_{11}$ and $P_{21}$. And the original and final state, as well as the frequency conditions corresponding to the transitions can also be found in TABLE~\ref{Table1} and TABLE~\ref{Table2}, respectively. These findings provide theoretical bases for controlling the transmission of light in this hybrid semiconductor optomechanical system via the mechanical resonator.

\section{CONCLUSIONS}\label{Sec.V}
In summary, we mainly study a hybrid system that consists of a cavity with a thin semiconducting membrane placed inside. Based on the properties of the semiconductor, which couples the excitons inside with phonons via deformation or piezoelectric potential, we build a fully coupled tripartite system. The physics of the system can be described in terms of the polaritons coupled to the mechanical resonator. We have determined analytically the eigenvalues of the corresponding Hamiltonian with generalized rotating-wave approximation. Besides, we study the dependence of emission spectrum for the polariton mode $B$ on the coupling strengths of the mechanical resonator to the cavity photon and the exciton. It is found that the spectrum of polariton displays series of peaks spaced with the frequency of the mechancial resonator $\omega_m$. And the visibility of higher-order phonon sidebands require larger photon- or exciton-phonon coupling strength. Moreover, we demonstrate the statistical properties of mode $B$ based on the equal-time second-order correlation. Compared to the balanced coupling case, i.e., the coupling strength between the mechanical resonator and cavity photons equals to that between the mechanical resonator and excitons, the equal-time second-order correlation exhibits more fine structure for the unbalanced photon- and exciton-phonon  coupling case. The frequency interval between neighbouring peaks and dips in the same series satisfies more complex condition than evenly spaced with $\omega_m$ or $\omega_m/2$ for the balanced coupling case. And the polariton changes from blockade into tunneling with increase of the exciton-phonon coupling strength.

%As an application reverse to previous works~\cite{Okamoto2011,Okamoto2011-PRB}, where fine energy structures of the excitons is detected from the spectrum of mechanical oscillation, 

Our study shows that the fine emission spectrum and blockade of the polariton actually shed light on the properties of mechanical resonator. This enables us to obtain the frequency of the mechanical resonator from the emission spectrum of the polariton. And the fine energy structures of the excitons can also be detected from the spectrum of the mechanical resonator as shown in Refs.~\cite{Okamoto2011,Okamoto2011-PRB}. This study provides a possible way to control light, sound and electric signals on an integrated platform and may also be applied to the quantum network for quantum entanglement between different quantum objects.

%and control the transmission of the light by modulating the detuning, i.e., the frequency difference between the pump field with the resonant frequency of polariton, and the photon- and exciton-phonon coupling strengths. 

\begin{acknowledgments}
Y.X.L. acknowledges the support of the National Basic Research Program of China 973
Program under Grant No.~2014CB921401, the Tsinghua University Initiative Scientific Research Program, and the Tsinghua National Laboratory for Information Science and Technology (TNList) Cross-discipline Foundation. Y. Zhang would like to acknowledge the support the NSF of China under Grant Nos.~11474189, 11674201. 
\end{acknowledgments}

\appendix  
\section{Eigenenergies and eigenstates in $\mathcal{N}=2$ subspace} 
In this Appendix, we provide explicit expressions for the eigenvalues in the $\mathcal{N}=2$ subspace. Similarly, in the zeroth-order approxiamtion, the Hamiltonian has the form 
\begin{eqnarray}
H_{3}^{\left( 0\right) }&=&\left( \omega _{A}-\omega _{B}\right) \cos \phi
J_{z}+\omega _{m}b^{\dagger }b-\frac{1}{\omega _{m}}{\left( 2\Omega +GJ_{z}\right) ^{2}}\nonumber\\
&&+\left( \omega _{A}-\omega _{B}\right) \sin \phi
J_{x}G_{0}\left( b^{\dagger }b\right). \label{H330}
\end{eqnarray}
For clearness, we leave out the constant term $\omega _{A}+\omega _{B}$. And the Hilbert space can be decomposed into different $n_{b}$ manifolds spanned by the angular momentum operator and oscillator basis of $\left\vert
1,-1,n_{b}\right\rangle ,$ $\left\vert 1,0,n_{b}\right\rangle $ and $\left\vert 1,1,n_{b}\right\rangle$. 
For $n_{b}$-th manifold, the Hamiltonian takes the form%
\[
H_{n_{b}}^{\left( 0\right) }=\left[ 
\begin{array}{ccc}
 e_{n_b}^{(1)}& \frac{\sqrt{2}}{2}B_{n_{b}} & 0
\\ 
\frac{\sqrt{2}}{2}B_{n_{b}} &e_{n_b}^{(2)} & \frac{\sqrt{2}}{2}B_{n_{b}} \\ 
0 & \frac{\sqrt{2}}{2}B_{n_{b}} & e_{n_b}^{(3)}
\end{array}%
\right] ,
\]%
with 
\begin{eqnarray}
e_{n_b}^{(1)}&=&\left( \omega _{B}-\omega _{A}\right) \cos \phi +n_{b}\omega _{m}-\frac{1}{\omega _{m}}{\left( 2\Omega -G\right) ^{2}},\\
e_{n_b}^{(2)}&=& n_{b}\omega _{m}-\frac{1}{\omega _{m}}{\left( 2\Omega \right)
^{2}},\\
e_{n_b}^{(3)}&=&\left( \omega _{A}-\omega _{B}\right) \cos
\phi +n_{b}\omega _{m}-\frac{1}{\omega _{m}}{\left( 2\Omega +G\right) ^{2}},\\
B_{n_{b}}&=&\left( \omega _{A}-\omega _{B}\right) \sin \phi
G_{0}\left( n_{b}\right).
\end{eqnarray}
The determinant of a matrix in this form gives the cubic equation $\lambda
^{3}+r\lambda +s=0$ and the eigenvalue $\varepsilon=\lambda +\frac{1}{3}\left(e_{n_b}^{(1)}+e_{n_b}^{(2)}+e_{n_b}^{(3)}\right)$.
Here 
\begin{eqnarray}
r &=&\frac{3ca-b^{2}}{3a^{2}},\\
s&=&\frac{2b^{3}-9abc+27a^{2}d}{%
27a^{3}} 
\end{eqnarray}
with 
\begin{eqnarray*}
a &=&1,\text{ \ \ }b=-\left( e_{n_b}^{(1)}+e_{n_b}^{(2)}+e_{n_b}^{(3)}\right), \\
c &=&e_{n_b}^{(1)}e_{n_b}^{(2)}+e_{n_b}^{(2)}e_{n_b}^{(3)}+e_{n_b}^{(3)}e_{n_b}^{(1)}-B_{n_{b}}^2,\\
d &=&-e_{n_b}^{(1)}e_{n_b}^{(3)}e_{n_b}^{(3)}+\frac{1}{2}(e_{n_b}^{(3)}-e_{n_b}^{(1)})B_{n_{b}}^2.\\
\end{eqnarray*}%
Then the corresponding eigenvalues $\varepsilon _{1,q,n_{b}}(q=1,2,3)$ are straightforwardly given by
\begin{eqnarray*}
\varepsilon _{1,1,n_{b}} &=&n_{b}\omega _{m}-\frac{1}{\omega _{m}}{[\left(2\Omega \right)
^{2}+\frac{2}{3}G^{2}]}+w\chi _{1,n_{b}}+w^{2}\chi _{2,n_{b}}, \\
\varepsilon _{1,2,n_{b}} &=&n_{b}\omega _{m}-\frac{1}{\omega _{m}}{[\left(2\Omega \right)
^{2}+\frac{2}{3}G^{2}]}+w^{2}\chi _{1,n_{b}}+w\chi _{2,n_{b}},\\
\varepsilon _{1,3,n_{b}} &=&n_{b}\omega _{m}-\frac{1}{\omega _{m}}{[\left(2\Omega \right)
^{2}+\frac{2}{3}G^{2}]}+\chi _{1,n_{b}}+\chi _{2,n_{b}}, 
\end{eqnarray*}
with 
\begin{eqnarray*}
w &=&\frac{1}{2}{(-1+\sqrt{3}i)},\\
\chi _{1,n_{b}} &=&\sqrt[3]{-\frac{s}{2}+\sqrt{\left( \frac{s}{2}\right)
^{2}+\left( \frac{r}{3}\right) ^{3}}}, \\
\chi _{2,n_{b}} &=&\sqrt[3]{-\frac{s}{2}-\sqrt{\left( \frac{s}{2}\right)
^{2}+\left( \frac{r}{3}\right) ^{3}}}.
\end{eqnarray*}
And the corresponding eigenfunctions
\begin{eqnarray*}
\left\vert \varepsilon _{1,q,n_{b}}\right\rangle &=&\frac{1}{\lambda
_{1,q,n_{b}}}\left( 
\begin{array}{c}
k_{1,q,n_{b}} \\ 
1 \\ 
f_{1,q,n_{b}}%
\end{array}%
\right),
\end{eqnarray*}
where 
\begin{eqnarray*}
k_{1,q,n_{b}} &=&\frac{\frac{\sqrt{2}}{2}B_{n_{b}}}{\varepsilon
_{1,q,n_{b}}-e_{n_b}^{(2)} },
\\
f_{1,q,n_{b}} &=&\frac{\frac{\sqrt{2}}{2}B_{n_{b}}}{\varepsilon
_{1,q,n_{b}}-e_{n_b}^{(3)} },
\\
\lambda _{1,q,n_{b}} &=&\sqrt{1+k_{1,q,n_{b}}^{2}+f_{1,q,n_{b}}^{2}}.
\end{eqnarray*}

As the first-order correction, we include the term $iJ_{y} \left[ F_{1}\left( b^{\dagger }b\right)b^{\dagger }-bF_{1}\left( b^{\dagger }b\right)\right]$. The Hamiltonian
now consists two parts:
\begin{eqnarray}
H_{3}^{\left( 1\right) }=H_{3,0}^{\left( 1\right) }+H_{3,1}^{\left( 1\right) },
\end{eqnarray}
with 
\begin{eqnarray}
H_{3,0}^{\left( 1\right) }=H_3^{(0)}-\left( \omega _{A}-\omega _{B}\right) \sin
\phi J_{x}\left[ G_{0}\left( b^{\dagger }b\right) -\beta \right],
\end{eqnarray}
and 
\begin{eqnarray}
H_{3,1}^{\left( 1\right) } &=&\left( \omega _{A}-\omega _{B}\right) \sin
\phi J_{x}\left[ G_{0}\left( b^{\dagger }b\right) -\beta \right]\nonumber\\
&&+i\left( \omega _{A}-\omega _{B}\right) \sin
\phi J_{y} \left[ F_{1}\left( b^{\dagger }b\right)b^{\dagger }-bF_{1}\left( b^{\dagger }b\right)\right].\nonumber\\
\end{eqnarray}
The angular momentum part in $H_{3,0}^{\left(1\right) }$ can be diagonalized in the basis of angular momentum in the $\mathcal{N}=2$ subspace, i.e., $\left\vert 1,-1\right\rangle$, $\left\vert 1,0\right\rangle$ and 
$\left\vert 1,1\right\rangle$, by a unitary matrix %
\begin{eqnarray}
U_4 &=&\left[ 
\begin{array}{ccc}
\frac{k_{1}}{\lambda _{1}} & \frac{1}{\lambda _{1}} & \frac{f_{1}}{\lambda
_{1}} \\ 
\frac{k_{2}}{\lambda _{2}} & \frac{1}{\lambda _{2}} & \frac{f_{2}}{\lambda
_{2}} \\ 
\frac{k_{3}}{\lambda _{3}} & \frac{1}{\lambda _{3}} & \frac{f_{3}}{\lambda
_{3}}%
\end{array}%
\right].
\end{eqnarray}
Then the total hamiltonian in the first-order approximation can be transformed into 
\begin{eqnarray}
\tilde{H}_{3,1}^{\left( 1\right) } &=&U_{4}H_{3,1}^{\left( 1\right) }U_{4}^{\dagger }
\nonumber\\
&=&L \left( \omega _{A}-\omega _{B}\right) \sin \phi \left[ G_{0}\left(
b^{\dagger }b\right) -\beta \right]  \nonumber\\
&&+M \left( \omega _{A}-\omega _{B}\right) \sin \phi \left[ F_{1}\left( b^{\dagger }b\right)b^{\dagger }-bF_{1}\left( b^{\dagger }b\right)\right].\nonumber\\
\end{eqnarray}
Here L is a symmetric matrix with
\begin{eqnarray}
L_{ii}&=&\frac{\sqrt{2}\left( k_{i}+f_{i}\right) }{\lambda _{i}^{2}},\\
L_{ij}&=&\frac{\sqrt{2}}{2}\frac{\left( k_{j}+f_{i}\right) +\left( k_{i}+f_{j}\right) }{\lambda
_{i}\lambda _{j}},
\end{eqnarray}
while M an antisymmetric one with
\begin{eqnarray}
M_{ij(i<j)}=\frac{\sqrt{2}}{2}\frac{\left( k_{i}+f_{j}\right) -\left( k_{j}+f_{i}\right) }{\lambda
_{i}\lambda _{j}}.
\end{eqnarray}

Once again, we neglect the static shift of the mechanical resonator which can be fully taken into account at the
expense of losing analytical expressions for the eigenenergies and
eigenvectors using Braak's method~\cite{Braak2011}. Besides, by neglecting the remote matrix elements $L_{1,3},L_{3,1},M_{1,3},M_{3,1}$ and the counter-rotating-wave
terms, i.e., $J_{+}b^{\dagger }+J_{-}b$, we can arrive at the total Hamiltonian  
\begin{widetext}
\begin{eqnarray}
H_{3}^{\rm{GRWA}} &=&\omega _{m}b^{\dagger }b+{\xi} _{1,N_{b}} \left\vert 1,-1\right\rangle \left\langle 1,-1\right\vert +{\xi} _{2,N_{b}} \left\vert 1,0\right\rangle
\left\langle 1,0\right\vert +{\xi} _{3,N_{b}} \left\vert 1,1\right\rangle
\left\langle 1,1\right\vert \nonumber \\
&&+M_{12}\left( \omega _{A}-\omega
_{B}\right) \sin \phi F_{1}\left( b^{\dagger }b\right) \left( b^{\dagger
}\left\vert 1,-1\right\rangle \left\langle 1,0\right\vert +b\left\vert
1,0\right\rangle \left\langle 1,-1\right\vert \right)\nonumber \\
&&+M_{23}\left( \omega _{A}-\omega
_{B}\right) \sin \phi F_{1}\left( b^{\dagger }b\right) \left( b^{\dagger
}\left\vert 1,0\right\rangle \left\langle 1,1\right\vert +b\left\vert
1,1\right\rangle \left\langle 1,0\right\vert \right) 
\end{eqnarray}
\end{widetext}
with
\begin{eqnarray}
{\xi} _{i,N_{b}}=\varepsilon _{i}+\left( \omega _{A}-\omega
_{B}\right) L_{ii}\sin\phi \left( G_{0}\left(b^{\dagger}b \right) -\beta \right).
\end{eqnarray}

The individual bosonic creation (annihilation) operator $b^{\dagger }\left(
b\right) $ also appears in the GRWA, so the transitions between states
belonging to different manifolds should be involved. In the basis of $%
\left\vert 1,-1,n_{b}+1\right\rangle ,\left\vert 1,0,n_{b}\right\rangle $
and $\left\vert 1,1,n_{b}-1\right\rangle $ $\left( n_{b}=1,2,\cdots \right)$, 
the Hamiltonian in the $n_b$-th block $H_{3,n_b}^{\rm{GRWA}}$ takes the following matrix form%
\begin{widetext}
\[
H_{3,n_b}^{\rm{GRWA}}=\left[ 
\begin{array}{ccc}
\left( n_{b}+1\right) \omega _{m}+\xi _{1,n_{b}+1} &\mathcal{P} & 0
\\ 
\mathcal{P}& n_{b}\omega _{m}+\xi _{2,n_{b}} &\mathcal{D} \\ 
0 & \mathcal{D} & \left( n_{b}-1\right) \omega _{m}+\xi
_{3,n_{b}-1}%
\end{array}%
\right], 
\]%
\end{widetext}
where 
\begin{eqnarray*}
\mathcal{P}&=& M_{12}\left( \omega _{A}-\omega _{B}\right) \sin \phi R_{n_{b},n_{b}+1}, \\
\mathcal{D}&=&M_{23}\left( \omega _{A}-\omega _{B}\right) \sin \phi R_{n_{b}-1,n_{b}},\\
\xi _{i,n_{b}}&=&\varepsilon _{i}+\left( \omega _{A}-\omega
_{B}\right) L_{ii}\sin\phi \left[ G_{0}\left( n_{b}\right) -\beta \right].
\end{eqnarray*}
Then the eigenvalues can be obtained as
\begin{eqnarray}
E_{1,1,n_{b}}^{\rm{GRWA}} &=&n_{b}\omega _{m}+\frac{1}{3}\left( \xi
_{1,n_{b}+1}+\xi _{2,n_{b}}+\xi _{3,n_{b}-1}\right)+\mathcal{Q}_1, \nonumber\\
E_{1,2,n_{b}}^{\rm{GRWA}} &=&n_{b}\omega _{m}+\frac{1}{3}\left( \xi
_{1,n_{b}+1}+\xi _{2,n_{b}}+\xi _{3,n_{b}-1}\right)+\mathcal{Q}_2,\nonumber\\
E_{1,3,n_{b}}^{\rm{GRWA}} &=&n_{b}\omega _{m}+\frac{1}{3}\left( \xi
_{1,n_{b}+1}+\xi _{2,n_{b}}+\xi _{3,n_{b}-1}\right) \nonumber\\
&&+\mu _{1,n_{b}}+\mu  
_{2,n_{b}}, \label{E1qnb}
\end{eqnarray}%
with
\begin{eqnarray*}
\mathcal{Q}_1&=&\min\left({w\mu_{1,n_{b}}+w^{2}\mu _{2,n_{b}}, w^{2}\mu
_{1,n_{b}}+w\mu _{2,n_{b}}}\right),\\
\mathcal{Q}_2&=&\max\left({w\mu_{1,n_{b}}+w^{2}\mu _{2,n_{b}}, w^{2}\mu
_{1,n_{b}}+w\mu _{2,n_{b}}}\right).
\end{eqnarray*}
Note that here we present the eigenvalues in ascending order. The parameter $\mu _{i,n_{b}}$ can be get with the same process as $\chi _{i,n_{b}}$ by solving the cubic equation. And the corresponding eigenstate has the form
\begin{eqnarray}
\left\vert \varphi _{1,q,n_{b}}^{\rm{GRWA}}\right\rangle &=&\frac{1}{\Lambda
_{1,q,n_{b}}}( K_{-1,q,n_{b}}\left\vert 1,-1,n_{b}+1\right\rangle
+\left\vert 1,0,n_{b}\right\rangle \nonumber\\
&&+F_{1,q,n_{b}}\left\vert
1,1,n_{b}-1\right\rangle ),
\end{eqnarray}
with
\begin{eqnarray*}
K_{1,q,n_{b}} &=&\frac{\mathcal{P}}{E_{1,q,n_{b}}^{\rm{GRWA}}-\left[ \left( n_{b}+1\right) \omega
_{m}+\xi _{1,n_{b}+1}\right] },\\
F_{1,q,n_{b}} &=&\frac{\mathcal{D}}{E_{1,q,n_{b}}^{\rm{GRWA}}-\left[ \left( n_{b}-1\right) \omega
_{m}+\xi _{3,n_{b}-1}\right] }, \\
\Lambda _{1,q,n_{b}} &=&\sqrt{%
1+K_{1,q,n_{b}}^{2}+F_{1,q,n_{b}}^{2}}. 
\end{eqnarray*}
There is a special case for $n_{b}=0$. In the basis of $\left\vert
1,-1,1\right\rangle $ and $\left\vert 1,0,0\right\rangle$, the Hamiltonian in this block can be written as 
\begin{eqnarray*}
H_{3,0}^{\rm{GRWA}}=\left[ 
\begin{array}{cc}
\omega _{m}+\xi _{1,1} & X \\ 
X & \xi _{2,0}%
\end{array}%
\right],
\end{eqnarray*}
with $X=M_{12}\left( \omega
_{A}-\omega _{B}\right) \sin \phi R_{0,1}$, and the eigenvalues are given by%
\begin{eqnarray}
E_{1,q,0}^{\rm{GRWA}}&=&\frac{1}{2}\left( \omega _{m}+\xi _{1,1}+\xi _{2,0}\right)\nonumber\\
&&\pm \frac{1}{2}\sqrt{\left( \omega _{m}+\xi _{1,1}-\xi _{2,0}\right) ^{2}+4X^{2}},
\end{eqnarray}
with $q=1,2$ denoting eigenenergies of the first and second excited states, respectively. The ground state is $\left\vert 1,-1,0\right\rangle $ with energy 
\begin{eqnarray}
\tilde{E}_{G}^{\rm{GRWA}}&=&\xi
_{-,0}=\varepsilon _{-}\nonumber\\
&=&-\frac{1}{\omega _{m}}{[\left( 2\Omega \right) ^{2}+\frac{2}{3}G^{2}]}+\chi _{1,0}+\chi _{2,0}.\label{EGN2}
\end{eqnarray}
The eigenfunctions to the original Hamiltonian has the form 
\begin{widetext}
\begin{eqnarray}
\left\vert \psi _{1,q,n_{b}}^{\rm{GRWA}}\right\rangle  &=&U_{1}^{\dagger
}U_{2}^{\dagger }U_4^{\dagger }\left\vert \varphi
_{1,q,n_{b}}^{\rm{GRWA}}\right\rangle  \nonumber\\
&=&U_{1}^{\dagger }\left[ 
\begin{array}{c}
\left\vert 1,-1\right\rangle \frac{1}{\Lambda _{1,q,n_{b}}}\left( \frac{k_{1}%
}{\lambda _{1}}K_{-1,q,n_{b}}\left\vert n_{b}+1\right\rangle _{1,-1}+\frac{%
k_{2}}{\lambda _{2}}\left\vert n_{b}\right\rangle _{1,-1}+\frac{k_{3}}{%
\lambda _{3}}F_{1,q,n_{b}}\left\vert n_{b}-1\right\rangle _{1,-1}\right) \\ 
+\left\vert 1,0\right\rangle \frac{1}{\Lambda _{1,q,n_{b}}}\left( \frac{%
K_{-1,q,n_{b}}}{\lambda _{1}}\left\vert n_{b}+1\right\rangle _{1,0}+\frac{1}{%
\lambda _{2}}\left\vert n_{b}\right\rangle _{1,0}+\frac{F_{1,q,n_{b}}}{%
\lambda _{3}}\left\vert n_{b}-1\right\rangle _{1,0}\right)  \\ 
+\left\vert 1,1\right\rangle \frac{1}{\Lambda _{1,q,n_{b}}}\left( \frac{f_{1}%
}{\lambda _{1}}K_{-1,q,n_{b}}\left\vert n_{b}+1\right\rangle _{1,1}+\frac{%
f_{2}}{\lambda _{2}}\left\vert n_{b}\right\rangle _{1,1}+\frac{f_{3}}{%
\lambda _{3}}F_{1,q,n_{b}}\left\vert n_{b}-1\right\rangle _{1,1}\right) 
\end{array}%
\right].  \label{wavefunction1qnb}
\end{eqnarray}
\end{widetext}
Here $n_b=1,2,3\cdots$ and $q=1,2,3$. On the other hand, for $n_b=0$, the first and second-excited eigenstates are labeled by 
\begin{widetext}
\begin{eqnarray}
\left\vert \psi _{1,q,0}^{\rm{GRWA}}\right\rangle  &=&U_{1}^{\dagger
}U_{2}^{\dagger }U_5^{\dagger }\left\vert \varphi _{1,q,0}^{\rm{GRWA}}\right\rangle 
\nonumber\\
&=&U_{1}^{\dagger }\frac{1}{\Lambda _{1,q,0}}\left[ 
\begin{array}{c}
\left\vert 1,-1\right\rangle \left( \frac{1}{\Lambda _{1,-,0}}\left\vert
1\right\rangle _{1,-1}+\frac{\Upsilon _{1,q,0}}{\Lambda _{1,+,0}}\left\vert
0\right\rangle _{1,-1}\right)  \\ 
+\left\vert 1,0\right\rangle \left( \frac{\Upsilon _{1,1,0}}{\Lambda _{1,1,0}%
}\left\vert 1\right\rangle _{1,0}+\frac{\Upsilon _{1,2,0}}{\Lambda _{1,2,0}}%
\Upsilon _{1,q,0}\left\vert 0\right\rangle _{1,0}\right) 
\end{array}%
\right],   \label{wavefunction1q0}
\end{eqnarray}
\end{widetext}
with $q=1,2$ and
\begin{eqnarray*}
U_5 &=&\left[ 
\begin{array}{cc}
\frac{1}{\Lambda _{1,1,0}} & \frac{\Upsilon _{1,1,0}}{\Lambda _{1,1,0}} \\ 
\frac{1}{\Lambda _{1,2,0}} & \frac{\Upsilon _{1,2,0}}{\Lambda _{1,2,0}}%
\end{array}%
\right], \\
\Upsilon _{1,q,0} &=&\frac{1}{X}[E_{1,q ,0}^{\rm{GRWA}}-\left( \omega _{m}+\xi
_{1,1}\right) ] ,\\
\Lambda _{1,q,0} &=&\sqrt{1+\Upsilon _{1,q,0}^{2}}.
\end{eqnarray*}%
The eigenfunction corresponding to the ground state is 
\begin{eqnarray}
\left\vert \tilde\psi _{G}^{\rm{GRWA}}\right\rangle  &=&U_{1}^{\dagger }U_{2}^{\dagger
}\left\vert 1,-1,0\right\rangle \nonumber\\
&=&\sin ^{2}\frac{\phi }{2}\left\vert 1,1\right\rangle \left\vert
0\right\rangle _{1,-1}+\frac{\sin \phi }{\sqrt{2}}\left\vert
1,0\right\rangle \left\vert 0\right\rangle _{1,-1}\nonumber\\
&&+\cos ^{2}\frac{\phi }{2}%
\left\vert 1,-1\right\rangle \left\vert 0\right\rangle _{1,-1}. \label{GroundN2}
\end{eqnarray}% 

\end{document}